\documentclass[trackchanges]{aastex702}
\usepackage{amsmath}

\newcommand{\radyn}{\texttt{RADYN}}
\newcommand{\arms}{\texttt{ARMS}}
\newcommand{\kglobal}{\texttt{kglobal}}

\newcommand{\radynfp}{\texttt{RADYN+FP}}
\newcommand{\fpc}{\texttt{FP}}

\newcommand{\kms}{km s$^{-1}$}

\makeatletter
\let\frontmatter@title@above=\relax
\makeatother
\begin{document}

\title{A New Framework for Modeling Solar Flares from MHD to Kinetic Processes}

\author[orcid=0000-0003-4227-6809,sname=Allred]{Joel C. Allred}
\affiliation{NASA Goddard Space Flight Center, Solar Physics Laboratory, Code 671, Greenbelt, MD, 20771, USA}
\email[show]{joel.c.allred@nasa.gov}  

\author[orcid=0000-0001-5316-914X,sname=Kerr]{Graham S. Kerr} 
\affiliation{SUPA School of Physics and Astronomy, University of Glasgow, Glasgow, G12 8QQ UK}
\affiliation{Department of Physics, Catholic University of America, 620 Michigan Avenue Northeast, Washington, DC 20064, USA}
\email{Graham.Kerr.2@glasgow.ac.uk}

\author[orcid=0000-0003-1439-4218,sname=Guidoni]{Silvina E. Guidoni}
\affiliation{Department of Physics, American University, 4400 Massachusetts Avenue, NW, Washington, DC 20016, USA}
\email{guidoni@american.edu}

\author[orcid=0000-0002-9493-4730,sname=Dahlin]{Joel T. Dahlin}
\affiliation{NASA Goddard Space Flight Center, Space Weather Laboratory, Code 674, Greenbelt, MD, 20771, USA}
\affiliation{Astronomy Department, University of Maryland, College Park, MD 20742, USA}
\email{joel.t.dahlin@nasa.gov}

\author[orcid=0000-0002-5435-3544,sname=Swisdak]{Marc Swisdak}
\affiliation{Institute for Research in Electronics and Applied Physics, University of Maryland, College Park, MD 20742, USA}
\email{swisdak@umd.edu}

\author[orcid=0000-0002-6975-5642,sname=Karpen]{Judith T. Karpen}
\affiliation{NASA Goddard Space Flight Center, Space Weather Laboratory, Code 674, Greenbelt, MD, 20771, USA}
\email{judith.t.karpen@nasa.gov}

\author[orcid=0000-0002-0934-4550,sname=Tenishev]{Valeriy Tenishev}
\affiliation{Heliophysics and Planetary Science Branch, NASA Marshall Space Flight Center, Huntsville, AL, USA}
\email{valeriy.m.tenishev@nasa.gov}

\begin{abstract}
Physical processes in solar flares span many orders of magnitude in spatiotemporal scales, so a single unified model accounting for all relevant processes remains computationally intractable. Instead, specialized codes have been developed, targeting specific domains that tackle specific aspects of the solar flare problem. Here, we present a comprehensive framework that links three models to capture flare initiation and energy release, the subsequent particle acceleration, the propagation and thermalization of those particles, and ultimately the heating of the lower solar atmosphere. We use the 2.5D \arms\ magnetohydrodynamics (MHD) code to model stressing of the magnetic field in a simulated active region, the formation of a current sheet, magnetic reconnection, and the evolution of newly formed flare loops. We use the conditions in the \arms' current sheet to initialize the \kglobal\ model, which predicts time-dependent nonthermal electron and proton distributions. Finally, we use \radynfp\ to inject these distributions into a newly reconnected \arms\ loop and compute subsequent particle transport, atmospheric heating, and radiative emissions. This is the first detailed presentation of a \radynfp\ simulation tracking nonthermal electrons and protons in the same loop. As well as discussing the interesting physics that takes place, we note how this results in white-light flare emission height consistent with observations. This work is a first step toward building a fully 3D flare modeling framework. 
\end{abstract}


\section{Introduction} 
\label{sec:intro}

The rapid release of energy through the reconnection of stressed magnetic fields in stellar atmospheres drives eruptive phenomena such as flares, coronal mass ejections, and jets. During flares, newly reconnected magnetic loops are rapidly heated by highly energized charged particles accelerated at the looptops during the reconnection process. These particles propagate down the legs of the loops, colliding with and heating the ambient plasma until they are eventually thermalized in the denser chromospheric footpoints. In addition, thermal conduction fronts arising from flare heating of the corona heat the foopoints, though propagate more slowly from the looptops than the beamed particles. The rapid footpoint heating produces an overpressure that drives ablation of the newly heated plasma up the loop legs, which dramatically brighten with emissions across the electromagnetic spectrum. At the same time, the top of the chromosphere is compressed. The bright footpoints of numerous flare loops, together spanning a common polarity inversion line, form the two-ribbon structures that characterize the base of flare arcades.  

The dominant flare energy release and transport mechanisms range from the restructuring of magnetic fields on the global scale to kinetic-scale particle acceleration and transport. The overall morphology of solar eruptions (flares and CMEs) is well represented by three-dimensional extensions of the CSHKP standard flare scenario \citep{1964NASSP..50..451C,1966Natur.211..695S,1974SoPh...34..323H,1976SoPh...50...85K,schmieder2015,janvier2017,shen2022,dahlin2025}, which posits that magnetic reconnection in a current sheet created between oppositely oriented magnetic fields \citep{2000JGR...105.2375L} rapidly energizes the plasma and accelerates particles in newly forming flare loops while overlying flux is reconfigured into a confined or erupting flux rope. Energy is transported into the chromospheric footpoints of the loops through thermal conduction and particle beams, forming flare ribbons and driving chromospheric evaporation. Collectively these loops constitute the flare arcade. Reconnection also generates Alfv\'en waves \citep[e.g., ][]{karpen1998,axford1999,yang2025}, which not only contribute to turbulence and transport energy away from the reconnection site but also can accelerate ions.  

Different approaches using single models have been applied to this multiscale problem. With high-performance computing resources widely available, new models can incorporate more sophisticated treatments of the underlying physics than their predecessors \citep[e.g., ][]{2019NatAs...3..160C,kong2019,2022ApJ...941L..22K,shen2022,2024A&A...684A.171D}. However, the relevant structures and physical mechanisms span some ten orders of magnitude in scale, so no single model can resolve the evolution of all relevant processes across such disparate sizes. Instead, much work has gone into developing models that account for specific processes and spatial scales. 

In the present study, we have capitalized on these developments by linking a global-scale magnetohydrodynamic (MHD) model of an eruption in an idealized active region with a hybrid (MHD+kinetic-scale) model of particle acceleration in the flare reconnection region and a radiative-hydrodynamic model of a flux tube in the resulting flare arcade into which the accelerated particles propagate. This approach enables us to combine self-consistent generation of the evolving eruptive-flare magnetic structure with state-of-the-art treatments of particles in the reconnection region that precipitate toward the footpoints of newly formed flare loops. 

Processes such as Fermi acceleration
\citep{2006Natur.443..553D,2021PhRvL.126m5101A}, a combination of Fermi and betatron acceleration \citep{2022ApJ...925..191G}, acceleration at a termination shock \citep{1986ApJ...305..553F, 2012ApJ...753...28G, 2015Sci...350.1238C}, or stochastic acceleration \citep{1996ApJ...461..445M,2004ApJ...610..550P} accelerate electrons and ions to high energies (tens to thousands of keV) during reconnection-driven eruptive events. For example, \citet{2016ApJ...820...60G} used an analytic model based on conservation of adiabatic invariants to estimate the energy gain of particles orbiting contracting magnetic islands formed by flare reconnection in an ultra-high resolution 2.5D MHD simulation of an eruptive flare/coronal mass ejection. In this case, the electron energy gain is dominated by betatron, with some contributions from Fermi acceleration \citep{Borovikov-2017-AJ}. Data from the same global simulation provided the background conditions for modeling pitch-angle and energy distribution function evolution as particles are sequentially energized in multiple islands, showing that the process can produce power laws and energy breaks \citep{2022ApJ...925..191G}. \citet{2024ApJ...965....6G} showed analytically how such sequential acceleration, but in two successive regions, can temporally evolve into a double power-law spectrum.
 
Energetic electrons are readily detected through X-ray bremsstrahlung from their collisions with ambient plasma, and their distributions are determined by inverting the X-ray spectra \citep{1971SoPh...18..489B}. Interestingly, ions may receive a larger fraction of the released magnetic energy than electrons \citep{2024PhPl...31f2901Y}, but determining their energy distributions remains challenging due to the lack of direct observational signatures for deca- and hecto-keV ions. Highly Doppler-shifted line emission from fast moving nonthermal atoms, created when precipitating energetic ions undergo charge exchange, offers a promising detection method. \citet{1976ApJ...208..618O} first modeled this effect for hydrogen Ly~$\alpha$. \citet{2023ApJ...945..118K} revisited this using models capable of better capturing the non-equilibrium ionisation of the chromosphere finding that while very transient, the effect is potentially observable with Solar Orbiter's SPICE instrument. Nevertheless, beyond isolated detections of MeV-range ions through $\gamma$-rays \citep[e.g.,][]{2006ApJ...644L..93H,2009ApJ...698L.152S}, direct observational detections of nonthermal ion distributions remain scarce. However, they may heat and compress the solar atmosphere in ways that are quite different from nonthermal electrons. For example, \citet{2024ApJ...960...80S} found that deca- to hecto-keV ions could be responsible for the acoustic disturbances in the photosphere known as sunquakes. 

Field-aligned 1D radiation-hydrodynamic models have been efficient tools for understanding how individual loops respond to flare heating \citep[for reviews and references see][]{2022FrASS...960856K,2023FrASS...960862K}. The low $\beta$ environment in the chromosphere and corona prevents cross-field conduction and transport, justifying the isolated-strand assumption. These models allow critical sub-meter resolution, capturing shocks, steep gradients, the narrow chromosphere-corona transition region, and the transport of nonthermal flare-accelerated particles. Incorporating comparable spatiotemporal precision and detailed non-local thermodynamic equilibrium chromospheric physics into multi-dimensional frameworks remains computationally prohibitive. Although modern multi-D simulations capture macroscopic flare evolution \citep{2019NatAs...3..160C,2020ApJ...896...97R,2023SoPh..298..134D,2023ApJ...955..105R,2024A&A...684A.171D,2026ApJ..1000..227C}, extreme computational costs generally preclude energetic particle integration, sophisticated chromospheric treatments, and broad parameter studies. Therefore, contemporary field-aligned codes remain indispensable for investigating flare energy-transport mechanisms. In this work, we use the \radynfp\ code to model the transport of nonthermal electron and proton distributions, predicted by \kglobal, injected into a new loop formed during flare reconnection predicted by the 2.5D \arms\ simulation used by \citet{2016ApJ...820...60G}, which had ultra-high resolution (8 levels of adaptive-mesh refinement).  

In Section~\ref{sec:codes} we describe the three fundamental codes whose outputs have been chained: the \arms\ MHD code (Section~\ref{sec:arms}), which has been configured to run in 2.5D spherical coordinates; the 2.5D \kglobal\ code, which predicts the flare-accelerated electron and proton distributions (Section~\ref{sec:kglobal}); and the 1D radiative hydrodynamic code \radynfp\ (Section~\ref{sec:radynfp}). In Section~\ref{sec:simulations}, we present the results of these chained simulations including descriptions of the transport of nonthermal particles, the hydrodynamic evolution, and emitted continuum radiation in our model flare loop. We present conclusions, including limitations of our current approach and plans for subsequent studies using this modeling framework, in Section \ref{sec:conc}.

\section{Modeling Codes} \label{sec:codes}
\subsection{\arms} \label{sec:arms}

We performed high-resolution simulations of an eruptive flare and coronal mass ejection with the well-tested Adaptively Refined Magnetohydrodynamics Solver (\arms) code \citep{antiochos1999,lynch2008,masson2013,2021ApJ...914...39L,wyper2024,dahlin2025}. \arms\  employs a finite-volume representation of the plasma, flux-corrected transport \citep{1991JCoPh..92..142D}, and full adaptive mesh refinement (AMR) with PARAMESH \citep{2000CoPhC.126..330M}. The use of a staggered grid ensures that the magnetic field remains divergence free to machine roundoff. Because the Lundquist number of the solar corona is at least 6 orders of magnitude larger than is reachable by any MHD model, we use numerical resistivity and high resolution at current sheets to attain Lundquist numbers large enough to enable plasmoid instability \citep[e.g., ][]{bhattacharjee2009, huang2017}) while minimizing diffusivity away from current sheets. 

\subsection{\kglobal}\label{sec:kglobal}

The \kglobal\ computational model combines MHD fluid and particle descriptions while eliminating the need to resolve kinetic scales \citep{drake19a,2024PhPl...31f2901Y,yin24b}.  \kglobal\ simulations include four distinct plasma species: fluid protons and fluid electrons (which collectively form an MHD-like backbone), and macro-particle protons and electrons that move through the fluid grid. The particles are treated in the guiding-center limit, which eliminates the need to resolve their respective Larmor radii. Simulations are performed in a domain with two meaningful spatial coordinates and one invariant coordinate (sometimes called 2.5D).  Particles move across the magnetic field with an $\mathbf{E}\times\mathbf{B}$ drift and along the field with their parallel velocities. The reconnecting component of the upstream magnetic field (along the $x$ direction) and the total proton density (the combined number density of particle and fluid protons) serve as normalization parameters by defining the Alfv\'en speed $C_{A0}$. Since kinetic scales are excluded by the model, lengths are normalized to an arbitrary macroscale $L_0$ and time scales to $\tau_A = L_0 / C_{A0}$.  Temperatures and particle energies are normalized to $m_p C_{A0}^2$, which means that the intrinsic heating of the plasma scales with this parameter. The values used for these normalizations are given in Section 3.2. 

\kglobal\ simulations of macroscopic magnetic reconnection produce nonthermal particles with spectra extending up to three orders of magnitude in energy \citep{arnold2019,2024PhPl...31f2901Y}.  These spectra can generally be fit by a sum of a Maxwellian and a $\kappa$ distribution, with the former providing the thermal bulk population and the latter the power-law tails of the distribution.  The guide field -- the component of the magnetic field perpendicular to the plane of reconnection -- has been found to be a key control parameter, with nonthermal particle production rapidly falling as the magnitude of the guide field approaches that of the reconnecting component \citep{dahlin14a,2021PhRvL.126m5101A}.

\subsection{\radynfp} \label{sec:radynfp}

We simulate radiative-hydrodynamic evolution of a model flare loop in response to the heating and momentum deposition resulting from electron and proton distributions injected at the looptop using \radyn\ \citep{1992ApJ...397L..59C,1995ApJ...440L..29C,1997ApJ...481..500C,1999ApJ...521..906A,2005ApJ...630..573A,2015ApJ...809..104A} together with the particle transport code Fokker-Planck \fpc\ \citep[hereafter, A20]{2020ApJ...902...16A}.  \radyn\ has been used extensively in previous works to study energy transport during solar flares \citep[e.g.,][and references therein]{2021ApJ...912..153K,2022ApJ...931...60A,2023A&A...673A.150C,2024ApJ...977..246E,2024ApJ...969..121K,2024ApJ...970...21K}. While this has primarily focused on non-thermal electron injection, other forms of energy transport have been experimented with also, including how radiative signatures might differ from electron-beam only scenario \citep[e.g.][]{2018ApJ...856..178P,2016ApJ...827..101K,2025ApJ...986...73L,2024ApJ...960...80S,2026NatAs..10..202K}. 

\radyn\ solves the 1D equations of hydrodynamics in a magnetic field-aligned geometry. These equations are coupled with non-equilibrium atomic level population equations (six levels of H, nine levels of He I and He II, and six levels of Ca II), on an adaptive grid. The radiative transfer equation is solved using a 1.5D plane-parallel approximation. \radyn\ has proven especially useful for flare modeling because of its ability to model optically thick, non-LTE lines and continua for many transitions that are important for energy transport in the chromosphere, where beamed particles deposit much of their energy. The transport of these particles is handled by \fpc, which models how they interact with the ambient plasma via dominant processes such as Coulomb collisions (CC), return currents (RC), and magnetic mirroring (MM), by solving a time-independent Fokker-Planck equation for arbitrary distributions of electrons and ions injected at the looptop. In modeling Coulomb collisions with the ambient plasma, \fpc\ assumes the target particles are thermal but makes no limiting assumptions of cold or warm targets.  \fpc's Coulomb collision operator uses general forms for the Rosenbluth potentials, which are applicable over large ranges of nonthermal particle energy and ambient plasma temperatures. The return current balances the beamed current in time and space, and is assumed to be comprised of drifting ambient electrons accelerated by microscale electric fields \citep{1990A&A...234..496V,2012ApJ...745...52H,2017ApJ...851...78A}. The plasma is Ohmic for time scales long compared to the collisional time scale \citep{1990A&A...234..496V}, which in flare loops is typically much less than the time scale of hydrodynamic evolution (e.g., approximately 40~ms in plasma at a temperature of 10~MK and a density of $10^{10}$ cm$^{-3}$), so we assume Ohm's law is valid and the return current Joule heats the ambient plasma. 

We have enhanced \fpc\ with the ability to model the injection of multiple species simultaneously, accounting for each species' interactions with the ambient plasma, how it contributes to the total current density, and hence the balancing return current. This addition is important for our present work, since \kglobal\ predicts both accelerated electron and proton distributions. Because the RC is oppositely directed to the total beam current, the RC electric field will typically decelerate beamed electrons but may accelerate beamed protons due to their opposite charge. The numerical method for solving the Fokker-Planck equation for an arbitrary species is described in detail in Section 2 of A20. Briefly, a matrix equation is constructed, which represents the Fokker-Planck differential equation with external forces (see also Equation~23 of A20),
\begin{equation}
A_a\left[f_a\right] f_a  = b_a,
\end{equation}
where $f_a$ is the distribution function for species, $a$, discretized on a phase-space grid in energy, pitch-angle and the field-aligned spatial dimension, $A_a[f_a]$ is the matrix operator, and $b_a$ is the right-hand side described in Section 2.6-7 of A20. The matrix equation is nonlinear, since the RC electric force depends on $f_a$, and is solved iteratively using the technique described in Section 2.6 of A20. At each iteration, the current density is reevaluated, and the RC components of $A_a[f_a]$ are recalculated. 

In the multiple-species version of FP, individual matrix equations are constructed for each species as described in Sections 2.2 – 2.5 of A20. In this work, we consider electrons and protons, but FP has been implemented with the ability to include arbitrary number of species, with arbitrary mass and charge. The set of matrix equations to be solved is,
\begin{equation}
\begin{split}
A_e\left[f_e,f_p\right] f_e & = b_e \\
A_p\left[f_e,f_p\right] f_p & = b_p
\end{split}
\end{equation} 
where $f_e$, $f_p$, $A_e[f_e,f_p]$,  $A_p[f_e,f_p]$, $b_e$, and $b_p$, are the distribution functions, matrix operators, and right-hand sides for electrons and protons, respectively. The matrix equations are coupled, since the RC electric force is a function of the current density, which now depends on $f_e$ and $f_p$. These are solved iteratively using the technique described in Section 2.6 of A20. Again, at each iteration, the current density is reevaluated and the RC components of $A_e[f_e,f_p]$ and $A_p[f_e,f_p]$ are updated. This technique couples multiple-species interactions through the RC but does not include Coulomb collisions between these species directly. We expect these collisions to be negligible, since the nonthermal particle densities are much smaller than the ambient density.

\section{Simulations} \label{sec:simulations}

\subsection{ARMS} \label{sec:ARMSsim}

The same initial configuration and energization used by \citet{2012ApJ...760...81K} with six refinement levels was simulated by \citet{2016ApJ...820...60G} and utilized by us here, with eight levels of refinement dictated by local gradients in $\vert$J/B$\vert$. The initial potential magnetic-field configuration corresponds to a global solar dipole plus an equatorial active region (represented by an octupolar spherical harmonic) in an spherically symmetric atmosphere in hydrostatic equilibrium. The radial component of the magnetic field at the solar surface $R_{\odot}$ 
is
\begin{equation}
B_r(R_{\odot},\theta) = B_{p} (2 \cos^3{\theta} - \cos{\theta}),
\label{e:ARMS_Br_init}
\end{equation}
where $B_{p}=10$ G is the field strength at the poles, and $\theta$ is the colatitute angle in radians. In the active region, the field achieves a maximum of approximately $2.5$ G, \textbf{which is approximately one to two orders of magnitude smaller than coronal active region fields \citep{2000ApJ...530..999M} and three orders of magnitude smaller than observed active region photospheric fields}. The initial temperature profile is $T=2 R_{\odot}/r$ MK with a base pressure of $P= 5.5 \times 10^{-2}$ dyn cm$^{-2}$. 

The following smoothly ramped slow photospheric shear flow, localized about the equatorial polarity inversion line, was imposed to form a filament channel and build up the free magnetic energy that powered the eruption

\begin{equation}
v_{\phi}(R_{\odot},\theta,t)=\left\{ 
\begin{array}{ll} 
-v_{\phi,0} \sin{\left(16\, \theta\right)}\left[1-\cos{(2 \pi t/10^5)} \right],  &  \  \frac{\pi}{2} -\frac{\pi}{16} \le \theta \le \frac{\pi}{2} +\frac{\pi}{16} \hbox{ and } 0 \hbox{ s} \le t \le 10^5\hbox{ s}, \vspace{0.2cm}   \\
0 &  \hbox{otherwise}. 
\end{array} \right.
\label{e:ARMS_vphi_init}
\end{equation}
where the maximum speed $v_{\phi,0} = 20$ \kms (see Figure 1 of \citet{2012ApJ...760...81K} for a sketch of the initial magnetic field and its evolution).
The 2.5D simulation presented here solves numerically the standard ideal magnetohydrodynamic equations in spherical coordinates, with translational invariance out of the plane of the calculation. Note that the modeled active region is approximately an order of magnitude larger in size than typical solar active regions, and the Alfv\'en speed in the inner core of the region is approximately four times smaller than in a typical active region, implying a time scale factor of $40$. 

\begin{figure}
    \epsscale{1.1}
     \plotone{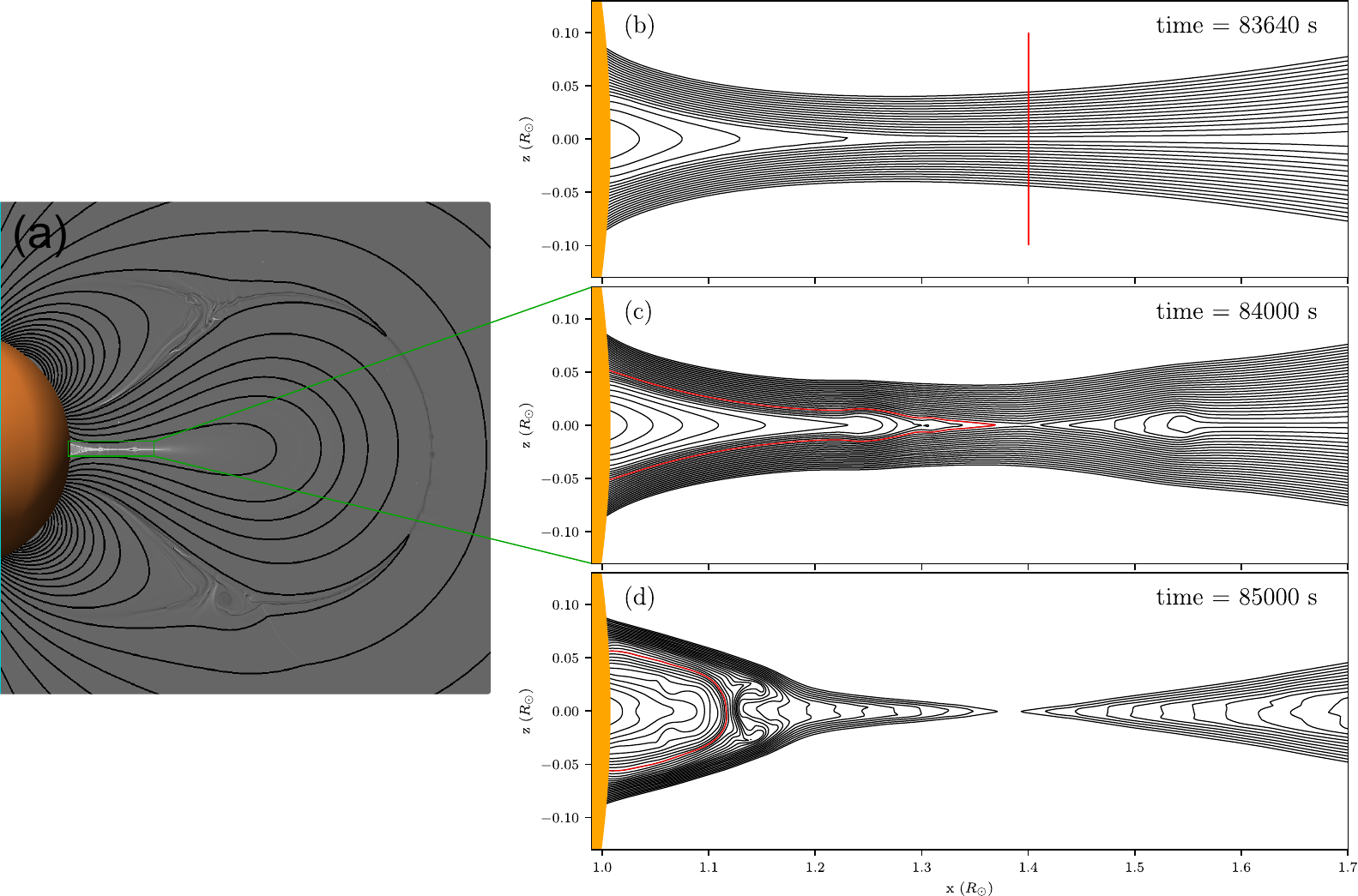}
     \caption{Selected flux surfaces (black) from the \arms\ simulation illustrating flare reconnection and the chosen loop for the \radynfp\ simulation. (a) Global configuration at $t=84,000$ s, just after flare reconnection onset. The orange sphere on the left is the solar surface, and the background grayscale plane indicates the normalized $\phi$-component (out of the plane) of the current density, $R_{\odot}J_{\phi}/c$, where $R_{\odot}$ is the solar radius and $c$ is the speed of light, saturated at a magnitude of $2$ G. (b)  Current sheet zoom-in just before reconnection onset (83,640 s). The plasma conditions upstream of the reconnecting current layer in the \kglobal\ simulations are based on those found at $x=1.4$ $R_{\odot}$ (vertical cut in red). (c) Flare arcade and current sheet zoom-in at the time of panel (a). 
     (d) Same region and flux surfaces as in (c), but at $t=85,000$s. The flux surface highlighted in red is shown immediately after its formation by magnetic reconnection (panel (c)), and then 1000~s later (panel (d)), by which time it has stopped shrinking and lies at the top of the flare arcade. Conditions from this latter time were used to construct the initial loop for the \radynfp\ simulation.}
     \label{fig:initloop}
\end{figure}

The slow buildup of magnetic energy in the active region is modeled by shearing the footpoints at sub-Alfv\'enic speeds in opposite directions within a narrow zone on either side of the polarity inversion line (PIL), forming a filament channel. The presence of a preexisting null in this configuration yields a current sheet above the stressed inner flux system, denoted the breakout current sheet, in which reconnection through plasmoid instability slowly removes strapping flux from inside the core, enabling the stressed flux to rise and form an elongated, thinning current sheet underneath – the flare current sheet. Plasma conditions in the vicinity of the flare current sheet at 83,640 s (Figure \ref{fig:initloop}(a)) were exported to provide initial conditions for the \kglobal\ simulations. At this time, the length of the flare current sheet is approximately $262$ Mm, equivalent to $26.2$ Mm when scaled to a typical active region size. Soon after reconnection commences in this current sheet, feedback between the breakout and flare reconnections accelerates the reconnection rate in both regions, causing explosive energy release at the main reconnection site in the flare current sheet and rapid reconfiguration of the associated magnetic field. The flare current sheet undergoes plasmoid-mediated reconnection, with plasmoids forming on either side of the main X-point. Plasmoids formed beneath the main null travel Sunward and merge with the flare arcade, while plasmoids formed above the main null travel outward and merge with the back of the CME. The impact of the reconnection outflow and later plasmoids on the growing flare arcade indents the top of the arcade (see Figure \ref{fig:initloop})

Figure \ref{fig:initloop}(a) shows the normalized $\phi$-component of the current density $R_{\odot}J_{\phi}/c$ in grayscale, where $R_{\odot}$ is the solar radius and $c$ is the speed of light, at $t$=84000 s. The elongated white structure centered at the equatorial plane corresponds to the flare arcade and current sheet, highlighted in panels (c) and (d). Starting at $t=84000$ s, soon after fast flare reconnection started, we analyzed $1000$ s of simulation time at a cadence of $20$ s (corresponding to $25$ s at a cadence of $0.5$ s for a typical active region.) Figure \ref{fig:initloop}(d) shows the flux surfaces of panel (c) at 85000 s, with the loop selected for the \radynfp\ simulation highlighted in red. 

We interpolated simulation data to a regular Cartesian grid, and, by tracing contours of the poloidal flux function, we followed flux surfaces of reconnected field lines on their descent and eventual arrival at the top of the flare arcade. The flux surface highlighted in red in Figure~\ref{fig:initloop}(c,d) was recently formed by reconnection and therefore was selected as the chosen loop for the \radynfp\ simulation. By 85,000~s this surface has stopped shrinking, and lies at the top of the flare arcade. Plasma variables (temperature, density, pressure, velocity and magnetic field components) at 85,000~s along this reconnected flux surface were exported to yield initial conditions for the \radynfp\ simulations. 

\subsection{\kglobal} \label{sec:kglobalsim}

\begin{figure}
    \epsscale{0.5}
    \plotone{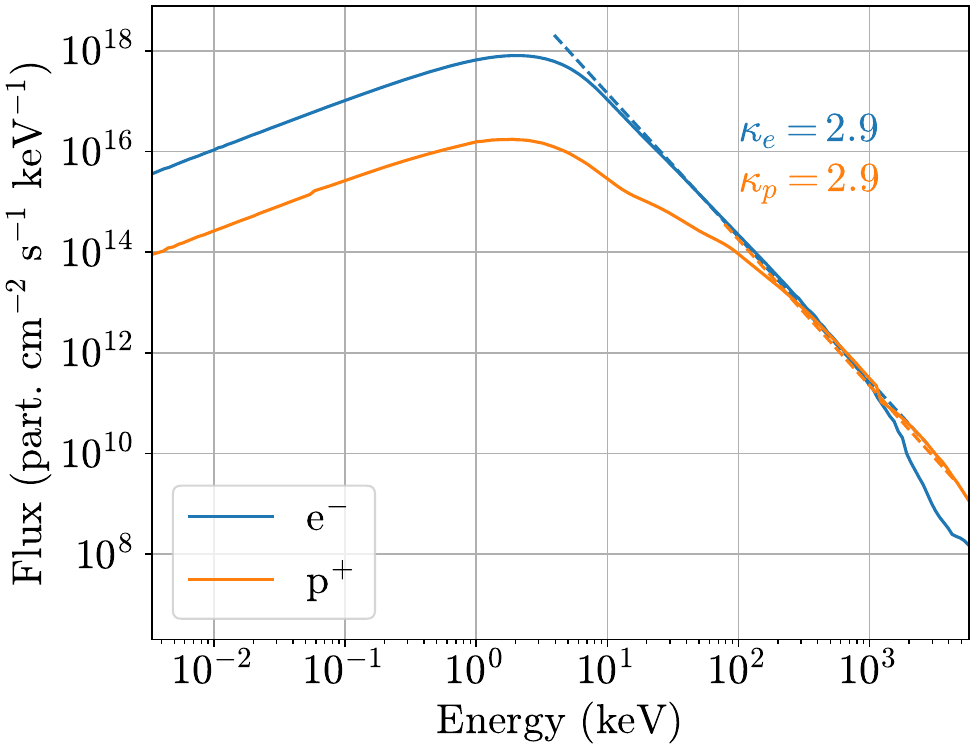}
    \caption{The flux distributions (solid lines) predicted by \kglobal\ for electrons (blue) and protons (orange). The dashed lines show power-law fits.}
    \label{fig:injecteddist}
\end{figure}

The plasma conditions upstream of the reconnecting current layer used to initiate the \kglobal\ simulations are based on those found by \arms\ at $t$=83,640 s (Figure~\ref{fig:initloop}(b)), giving an asymptotic $\beta$ of $\approx 0.1$. While the ratio of the guide field to the reconnecting field varies with space and time in the \arms\ simulation, we have chosen the representative value 0.3 for the \kglobal\ run.
The computational domain does not allow variations in the out-of-plane (guide field) direction and employs fully periodic boundary conditions.  Unlike the \arms\ simulation, here the focus is not on simulating a realistic flare geometry but instead on accurately describing the nonthermal spectra of particles produced during macroscale reconnection. The reconnecting component of the upstream magnetic field (along the $x$ direction) and the total proton density (the combined number density of particle and fluid protons) serve as normalization parameters by defining the Alfv\'en speed, $C_{A0} = 1733$~\kms, where $C_{A0}$ is taken from the plasma conditions upstream of the reconnecting layer in \arms\ and was multiplied by a factor of 4 to scale to typical active-region values as noted above. Since kinetic scales are excluded by the model, lengths are normalized to a macroscale $L_0 = 26.2/\pi = 8.3$~Mm, where 26.2~Mm is the scaled length of the \arms\ current sheet (Section 3.1). Then the time scale normalization yields $\tau_A = L_0 / C_{A0} = 4.8$~s. Temperatures and particle energies are normalized to $m_p C_{A0}^2 = 31.3$~keV, which means that the intrinsic heating of the plasma scales with this parameter. The simulation was evolved until both the proton and electron energy spectra reached a steady state ($\approx 12 \tau_A$), but most of the particle acceleration occurred within $5 \tau_A = 24$~s. The resulting spectra of electrons and protons was fit to a sum of a Maxwellian and a $\kappa$ distribution \citep{2021PhRvL.126m5101A,yin24b}. The spectral index of the $\kappa$ distribution, a key component determining the degree of nonthermal energization, varies significantly with guide field.  Here both the electrons and protons exhibit power laws with slope $\kappa \approx 2.9$, which is somewhat harder than typical for solar flares. The particle flux normalization is chosen so that the injected electron energy flux is initially $5 \times 10^{10}$ erg~cm$^{-2}$~s$^{-1}$, which corresponds to moderately strong flare heating. The final particle flux distributions are plotted as functions of energy in Figure~\ref{fig:injecteddist}.

\subsection{\radynfp} \label{sec:radynsim}
\begin{figure}
    \epsscale{0.7}
     \plotone{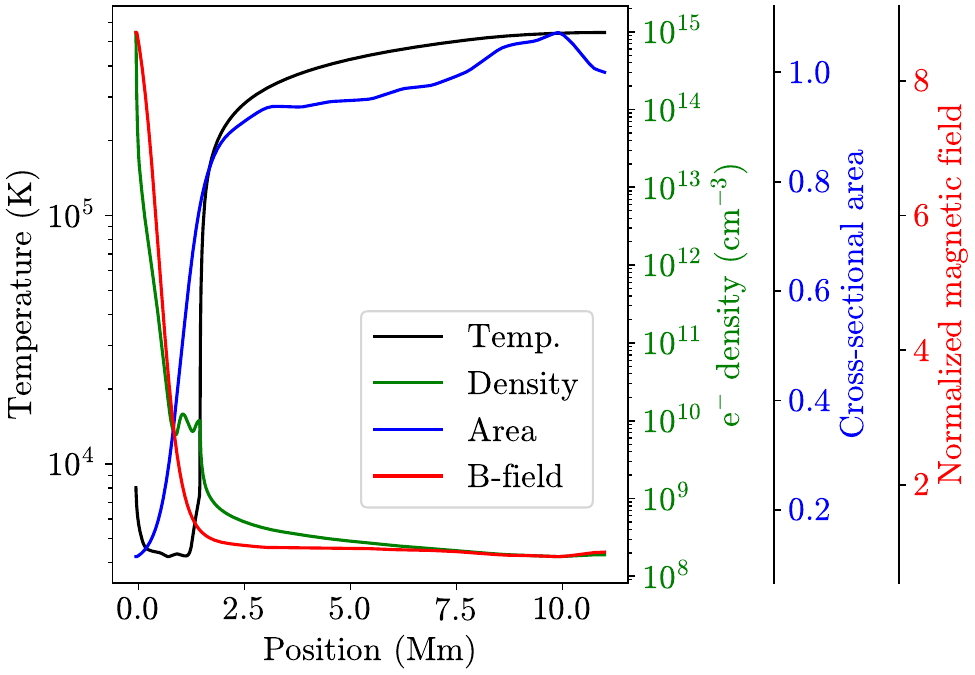}
     \caption{Temperature (black), electron density (green), normalized (to looptop) cross-sectional area (blue), and normalized (to looptop) magnitude of the magnetic flux density (red) as functions of position from the photosphere of the initial loop. $t=0$~s is the time when the particle distribution calculated by \kglobal\ begins injection into the \radynfp\ loop.}
     \label{fig:initloop_radyn}
\end{figure}

As noted in Section \ref{sec:ARMSsim}, plasma and loop conditions inputted to the \radynfp\ simulations are taken from \arms\ at $t = 85,000$~s, when the loop has fully retracted. The 1000~s interval between this time and reconnection onset corresponds to 25~s when scaled to typical active-region properties (Section 3.1). This time interval is comparable to $5 \tau_A$ (Section 3.2). \textbf{During the loop retraction, the coronal density and magnetic-field magnitude increase by factors of $\sim 4$ and $\sim 2$, respectively, resulting in a relatively constant Alfv\'en speed.} Similarly, to scale to typical active-region sizes when creating the initial pre-flare loop for \radynfp, the \arms\ loop length is scaled by a factor of 0.1, resulting in loop half-length of 11~Mm. \radyn's loop cross-sectional area is allowed to vary along the loop axis, and is assumed to be inversely proportional to the magnetic field magnitude (following \arms), so as to conserve magnetic flux. Because \arms\ simulations only include coronal plasma, we add \radyn's chromospheric and photospheric stratifications at the base of the loop model. We implement a converging magnetic field in the chromosphere using the technique described in \citet{2012ApJ...752....4B} and \citet{2023ApJ...943L..23K}. This converging field will be an important source of magnetic mirroring in the flare simulation described in the next section. Finally, we run \radyn, maintaining the coronal density, and with a small amount of coronal heating but without any other external sources of energy to counteract radiative losses, until the loop reaches a state of hydrostatic equilibrium, naturally forming a transition region. This steady state becomes the initial state for the flare simulation. Its coronal temperature is somewhat cooler than the \arms\ loop, but once the flare-accelerated particles are injected into the loop the coronal temperature very rapidly rises, and the relatively small initial temperature difference is negligible. The density, temperature, loop cross-sectional area, and normalized magnetic flux density magnitude ($|B|$) stratification are shown in Figure~\ref{fig:initloop_radyn}. Here the $B$-field is normalized to the value at the looptop, but the normalization is not critical for \radynfp\ since magnetic mirroring depends on $B$-field ratios.

\begin{figure}
    \epsscale{0.5}
    \plotone{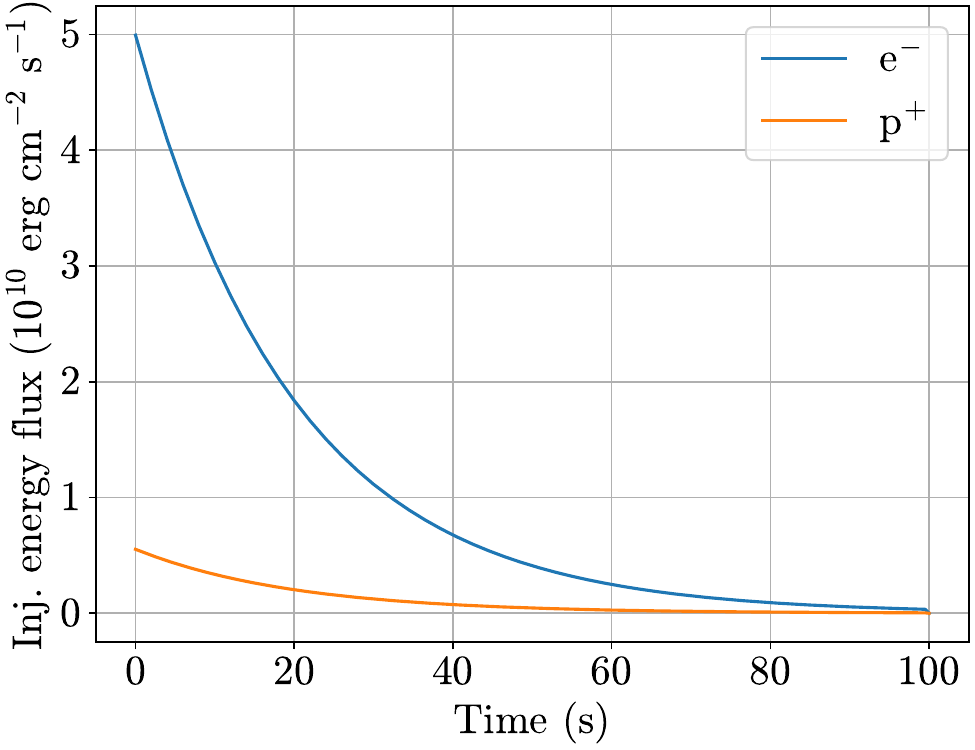}
    \caption{The injected energy fluxes in electrons (blue) and protons (orange) as functions of time.}
    \label{fig:injectedeflux}
\end{figure}

The accelerated particles predicted by \kglobal\ are injected into this \radynfp\ loop model. The \kglobal\ relaxation time to a steady state is fast compared to the injection time, so we make a ``leaky-box'' assumption that accelerated particles are injected into flare loops as they diffuse out of the acceleration region at a rate proportional to their speed and with isotropic pitch-angle distributions. The particles remaining in the acceleration region quickly re-equilibrate to the steady state. The electrons escape faster than the protons, carrying approximately 90\% of the injected energy flux even though, in the acceleration region at any given moment, the proton distribution is more energetic than the electron distribution. Charge neutrality is maintained by the return current, which moves drifting ambient electrons into the acceleration region. Thus in the leaky-box approximation, the energy dependence of the injected flux is steady in time but the flux normalization is a decaying exponential, with a characteristic time $\tau_{i}$.  Quasi-periodic pulsations (QPP) are potentially indicators of individual reconnection bursts \citep{2026NatAs..10...54A}. The intervals between QPPs are on the order of seconds to hundreds of seconds, with a typical time of about 20~s \citep{2020ApJ...895...50H}. We have chosen $\tau_{i} = 20$~s, making the assumption that the decay time is on a similar time scale as the reconnection bursts. The injected energy fluxes in electrons and protons as functions of time are plotted in Figure~\ref{fig:injectedeflux}. The distributions evolve over the course of the transport in response to interactions with the ambient plasma through Coulomb collisions, return currents, and magnetic mirroring. Energy lost by the distributions is assumed to be gained as heat by the ambient plasma. Thus the impulsive injection of these particles into flare loops drives rapid heating.  

\begin{figure*}[htb!]
    \epsscale{1.15}
    \plotone{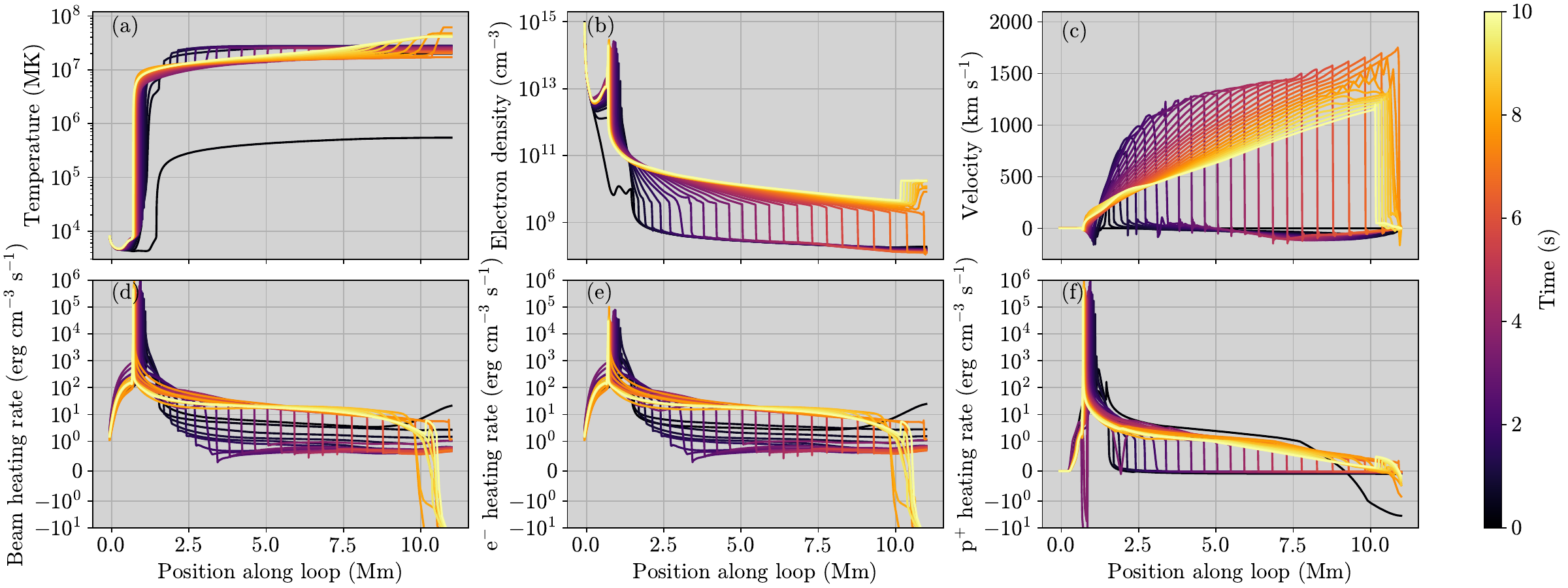}
    \caption{The first 10~s of temporal evolution of the (a) temperature, (b) electron density, (c) velocity (upward is positive), (d) summed electron and proton beam heating rates, (e) electron beam heating rates, and (f) proton beam heating rates as functions of position along the loop measured from the photosphere. Times are indicated by the line's color.}
    \label{fig:evol10s}
\end{figure*}

\begin{figure*}[htb!]
    \epsscale{1.15}
    \plotone{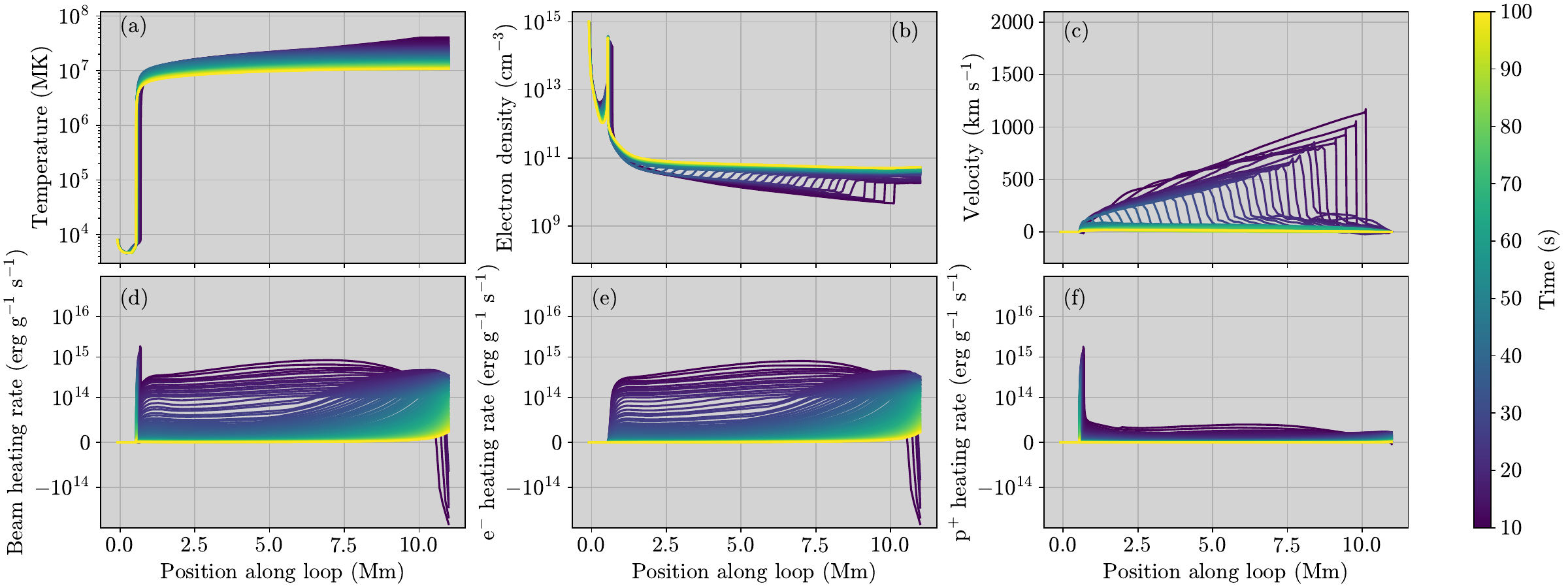}
    \caption{Same as Figure~\ref{fig:evol10s} but for temporal evolution in range $t = 10\textrm{--}100$~s.}
    \label{fig:evol100s}
\end{figure*}

The temporal evolution of the flare loop's temperature, electron density, beam heating rate (summed electrons and protons), and velocity, from 0--10~s and from 10--100~s is shown in Figures~\ref{fig:evol10s}~\&~\ref{fig:evol100s}, respectively. In the first time steps, the nonthermal electrons propagate through an initially cool, low density corona. In these conditions, energy losses in the looptop due to the return current electric field dominate over collisional losses. The Ohmic heating flash heats the looptop, resulting in an apex temperature of 24 MK, within $t = 0.5$~s. In this short time interval the RC electric field at the looptop exceeds the Dreicer field \citep{1959PhRv..115..238D}, meaning that a significant fraction of the return current will be carried by collisionless runaway electrons \citep{2021ApJ...917...74A}. This effect is not yet included in the \fpc\ transport code, which assumes a drifting Ohmic return current. Therefore, \fpc\ overestimates the plasma resistivity in this short time interval. However, the rapidly rising coronal temperature results in a corresponding drop in plasma resistivity ($\propto T^{-3/2}$), and hence the RC electric field. By $t = 0.5$~s the RC field in the coronal portion of the loop is relatively small and most of the beamed electrons impact the lower atmosphere. Very little material has flowed into the corona over such a short time interval so the emission measure remains small, but even this relatively small amount of Ohmic heating is able to further increase the coronal temperature, which reaches a peak of nearly 28~MK at 1.4~s. Interestingly, during the first seconds, the strong RC electric field accelerates the injected nonthermal protons, increasing their energies by approximately 2~keV over the course of their transport. Thus they have a net cooling effect near the looptop over this time interval. 

A fast moving evaporation flow brings material up from the lower atmosphere and reaches the loop apex at $t = 7.3$~s. Since the initial coronal density is quite low ($n_e = 10^{8}$~cm$^{-3}$), the upflow speed is at first very fast, peaking at $1.7$ Mm~s$^{-1}$, but levels off quickly as the coronal density increases, and as the initial upflow reflects off the loop apex (mimicking colliding flows from both footpoints of the symmetric loop), falling to values of $1.1$~Mm~s$^{-1}$ (apex $n_e = 2 \times 10^{10}$~cm$^{-3}$; $T = 40$~MK) at $t = 10$~s, 630 \kms (apex $n_e = 3 \times 10^{10}$~cm$^{-3}$; $T = 27$~MK) at $t = 20$~s, and 160 km~s$^{-1}$ (apex $n_e = 4 \times 10^{10}$~cm$^{-3}$; T = $18$~MK) at $t = 40$~s. 

\subsubsection{Nonthermal Particle Transport}

\begin{figure*}
\begin{interactive}{animation}{flxspec.mp4}
\plotone{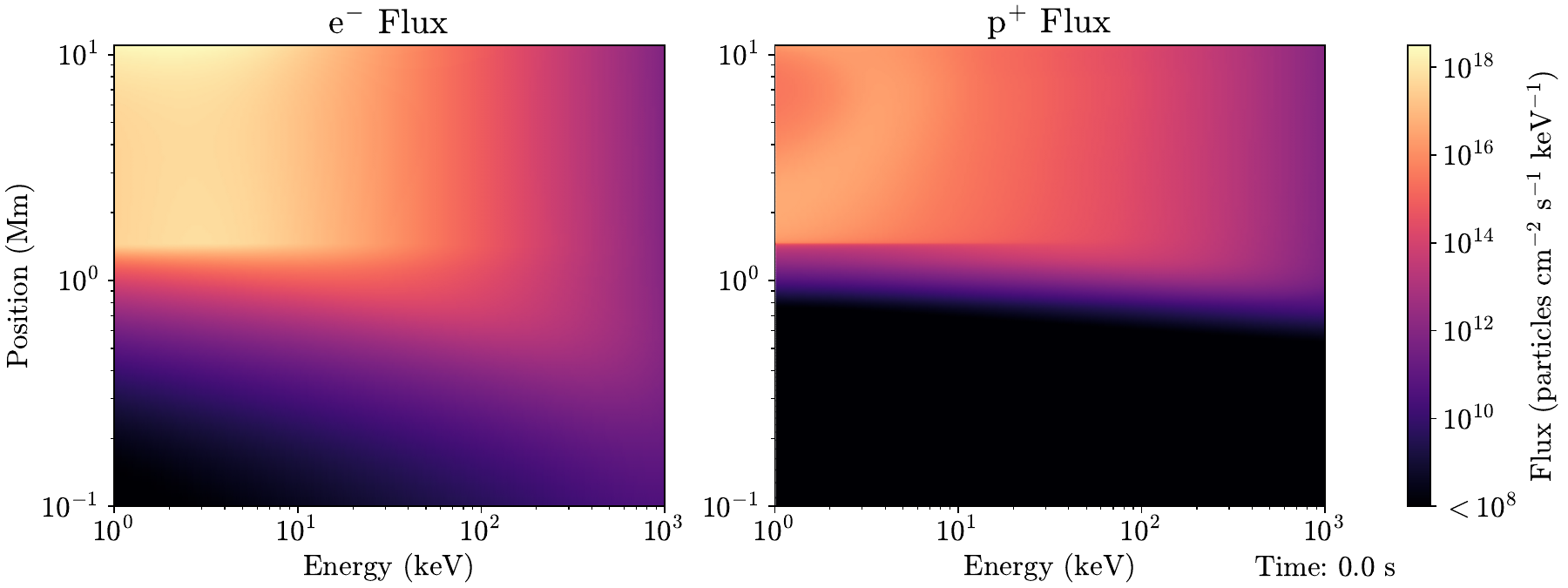}
\plotone{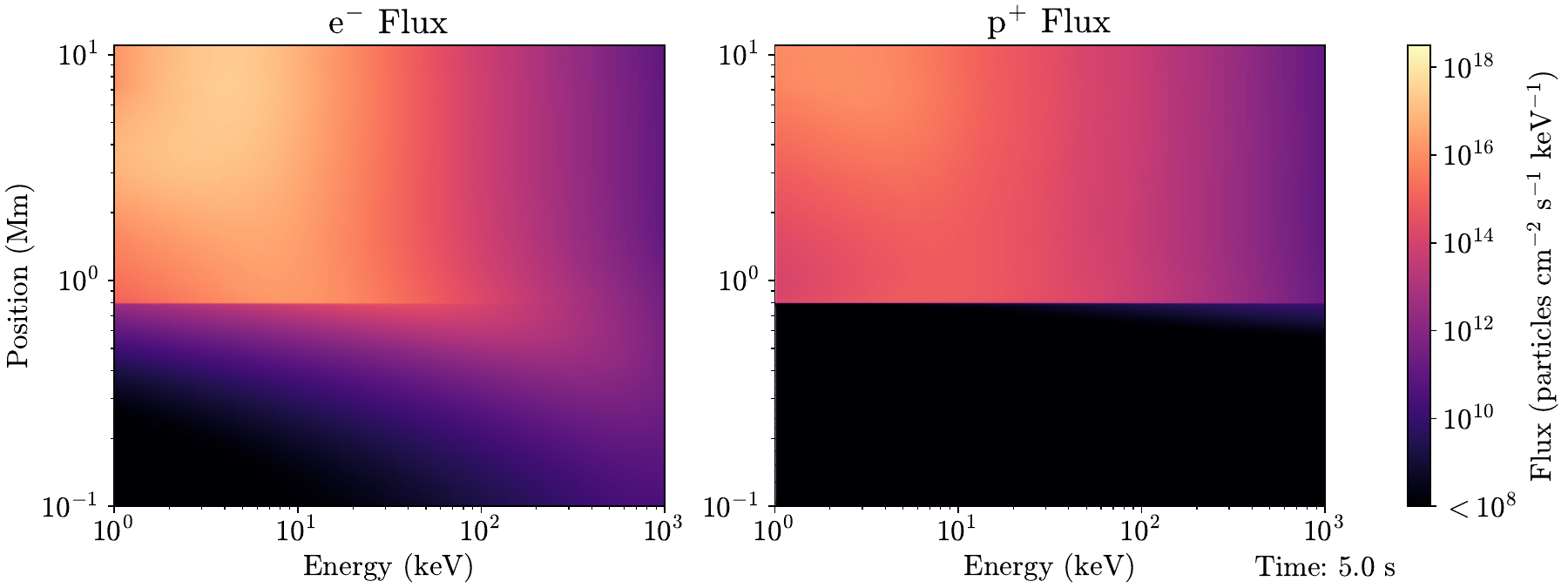}
\plotone{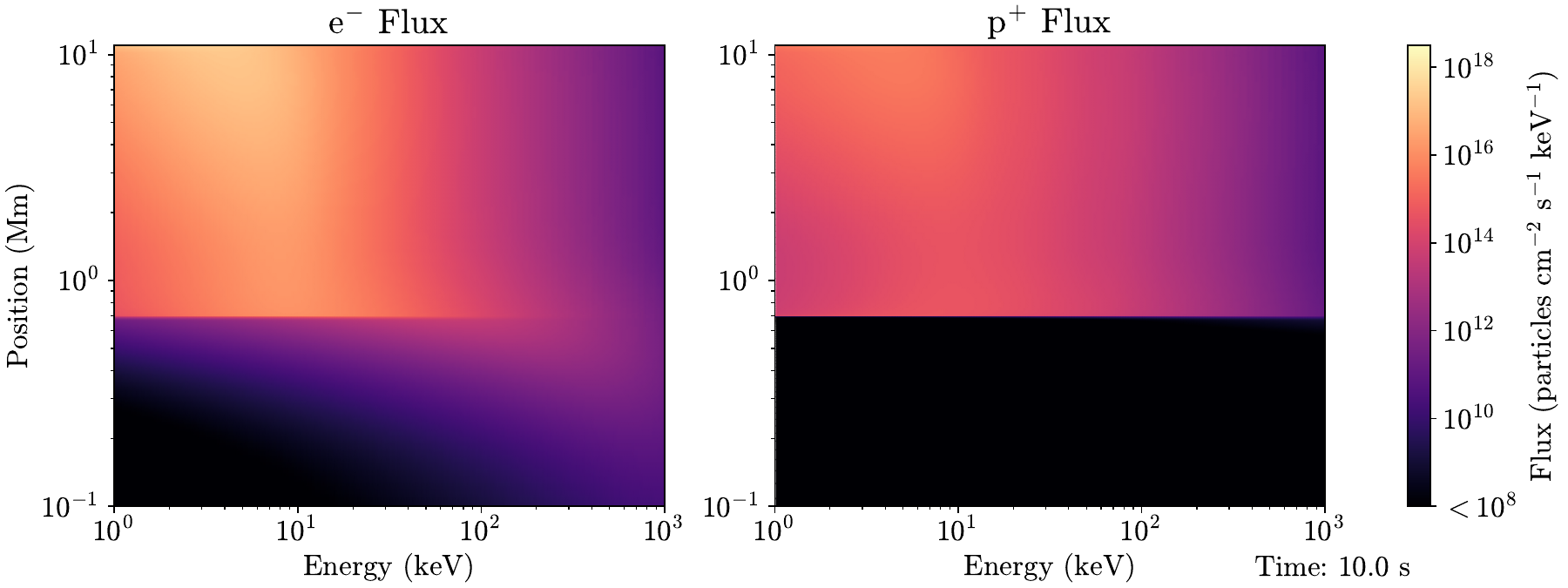}
\end{interactive}
\caption{Electron (left column) and proton (right column) flux distributions as functions of position (note log scale) and energy at $t=0$~s (top row), $t=5$~s (middle row), and $t=10$~s (bottom row). An animated version of this plot, showing the temporal evolution from $t=0$--99~s, is available online. The animation shows that initially, low energy electrons and protons penetrate to $\sim1.5$ Mm. As the simulation evolves the penetration depth is pushed lower to $\sim600$~km. High energy electrons penetrate much deeper ($<100$~km). Most of the protons experience warm-target collisions in the corona, which are ineffective at slowing them. Once they reach the transition region, the ambient electrons are cold-targets, effectively stopping them. This is evident in the animation as the sharp cutoff in flux at the transition region for all proton energies.}
\label{fig:flxspec}
\end{figure*}

The injected particle fluxes and their transport through the flare loop are the energy and momentum sources driving the flare dynamics. Figure~\ref{fig:flxspec} and associated movie show the temporal evolution of the electron and proton flux distributions as functions of position (measured from the photosphere) and energy. As mentioned above, after the first time steps the corona is sufficiently hot that return current losses there are small and Coulomb collisions dominate the energy losses. A small negative (upward-directed) electron flux -- visible as a small dark patch in the animation -- occurs near the looptop for low energy electrons ($< 2$~keV) just before and during the time when the initial chromospheric evaporation reaches the looptop ($t = 6.0\text{--}8.0$~s). These electrons have energies that are comparable to the ambient thermal value and are undergoing thermalization, so they have a nearly isotropic pitch-angle distribution; however, a relatively small magnetic mirroring force reflects some electrons, resulting in a small net negative flux. Once sufficient chromospheric evaporation has reached the looptop ($t > 8.0$~s), the density is large, and these low energy electrons are thermalized before the magnetic mirroring can reflect them. It is evident that higher energy nonthermal electrons penetrate deeper, as expected in cold-target collisions. However, this not true for the nonthermal protons, since they undergo warm-target collisions through much of their coronal transport.

Warm-target collisions are those in which the thermal speed of the ambient electrons exceeds that of the nonthermal particles \citep{1986ApJ...309..409T}. That is $x_e\ll1$, where $x_e$ is the square of the ratio of the nonthermal speed to thermal electron speed \citep[see Equation~8 of ][]{2020ApJ...902...16A}. For small $x_e$, the dynamic friction component of Coulomb collisions \citep[$F_c$ in Equation~9 of ][]{2020ApJ...902...16A} is also small, making these collisions ineffective at slowing nonthermal particles. Nonthermal protons interacting with, for example, 10~MK ambient electrons experience warm-target collisions for energies $\ll1.6$~MeV, i.e., nearly all nonthermal protons in the injected distribution. Thus, these protons will travel through the coronal portion of the loop with minimal energy loss, until they move into the lower transition region and chromosphere where the ambient electrons are cold-targets. Cold-target collisions are effective at stopping the nonthermal protons, so they deposit their energy and downward momentum, pushing the transition region lower. Interestingly, since the nonthermal protons heat this layer, the ambient electrons enter the warm-target regime, meaning newly impacting protons will penetrate to ever deeper layers. 

\begin{figure}
   \epsscale{0.5}
   \plotone{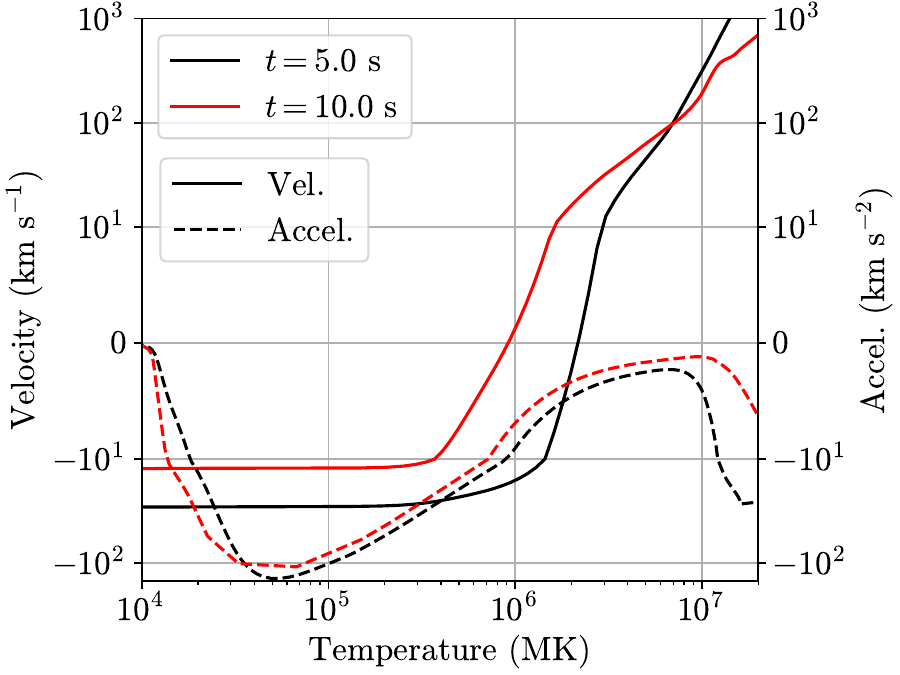}
   \caption{Bulk velocity (solid lines) and acceleration from beam impact (dashed lines) as functions of temperature in the loop at $t = 5$~s (black lines) and $t = 10$~s (red lines). Upward velocity is positive.}
   \label{fig:momvel}
\end{figure}

The beam of nonthermal protons carries momentum into the transition region pushing it downward. This is illustrated in Figure~\ref{fig:momvel} which shows the bulk acceleration produced by impacting nonthermal protons and the bulk velocity at $t=5$~s and $t = 10$~s. As is typically observed \citep[e.g.,][]{2009ApJ...699..968M,2011ApJ...740...70M}, hotter material moves upward with high speeds (chromospheric evaporation) and cooler material moves downward (chromospheric condensation). The chromospheric condensation results from the momentum transfer of the impacting nonthermal particles and from the pressure gradient produced as they heat the plasma \citep{1989ApJ...346.1019F}, but after the first seconds the pressure gradient dominates. The crossover from upward- to downward-directed flows occurs at 2~MK and about 1~MK at $t=5$~s and $t=10$~s, respectively. The downward-directed acceleration from impacting particles is more than 100~km s$^{-2}$ in the lower transition region but the force is toward increasingly dense plasma, and the velocity plateaus at 30 \kms\ and 12 \kms\ at $t=5$~s and $t=10$~s, respectively.

\begin{figure*}
    \plotone{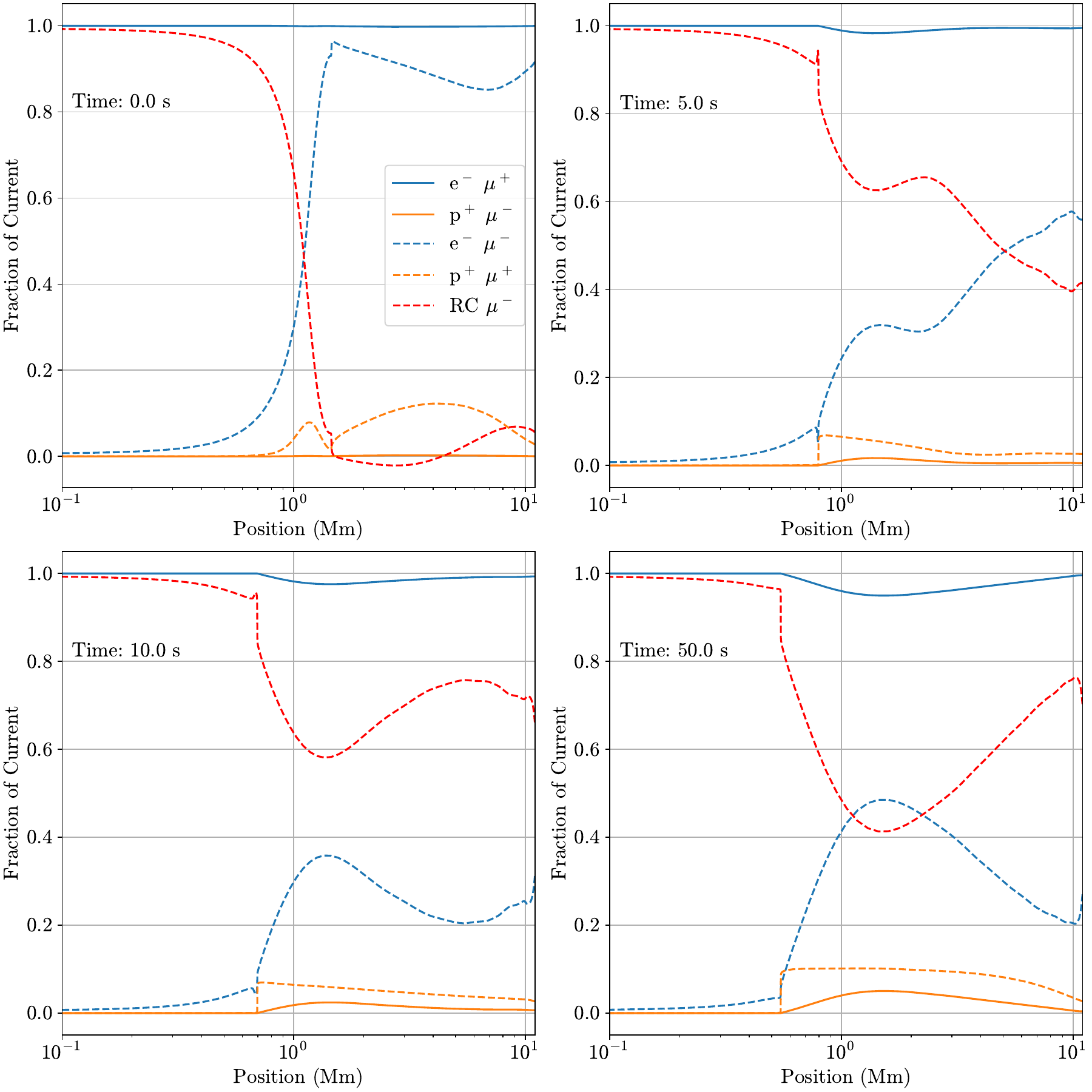}
    \caption{Fraction of total currents $J_{-}$ (solid lines) and $J_{+}$ (dashed lines) for nonthermal electrons (blue lines), nonthermal protons (orange lines) and drifting ambient electrons (red lines). At $t=0$~s (top left), $t=5$~s (top right), $t=10$~s (bottom left), $t=50$~s (bottom right).
    \label{fig:rcfrac}}
\end{figure*}

\begin{figure*}
    \begin{interactive}{animation}{testpart_e_20keV_th3.mp4}
    \epsscale{0.9}
    \plotone{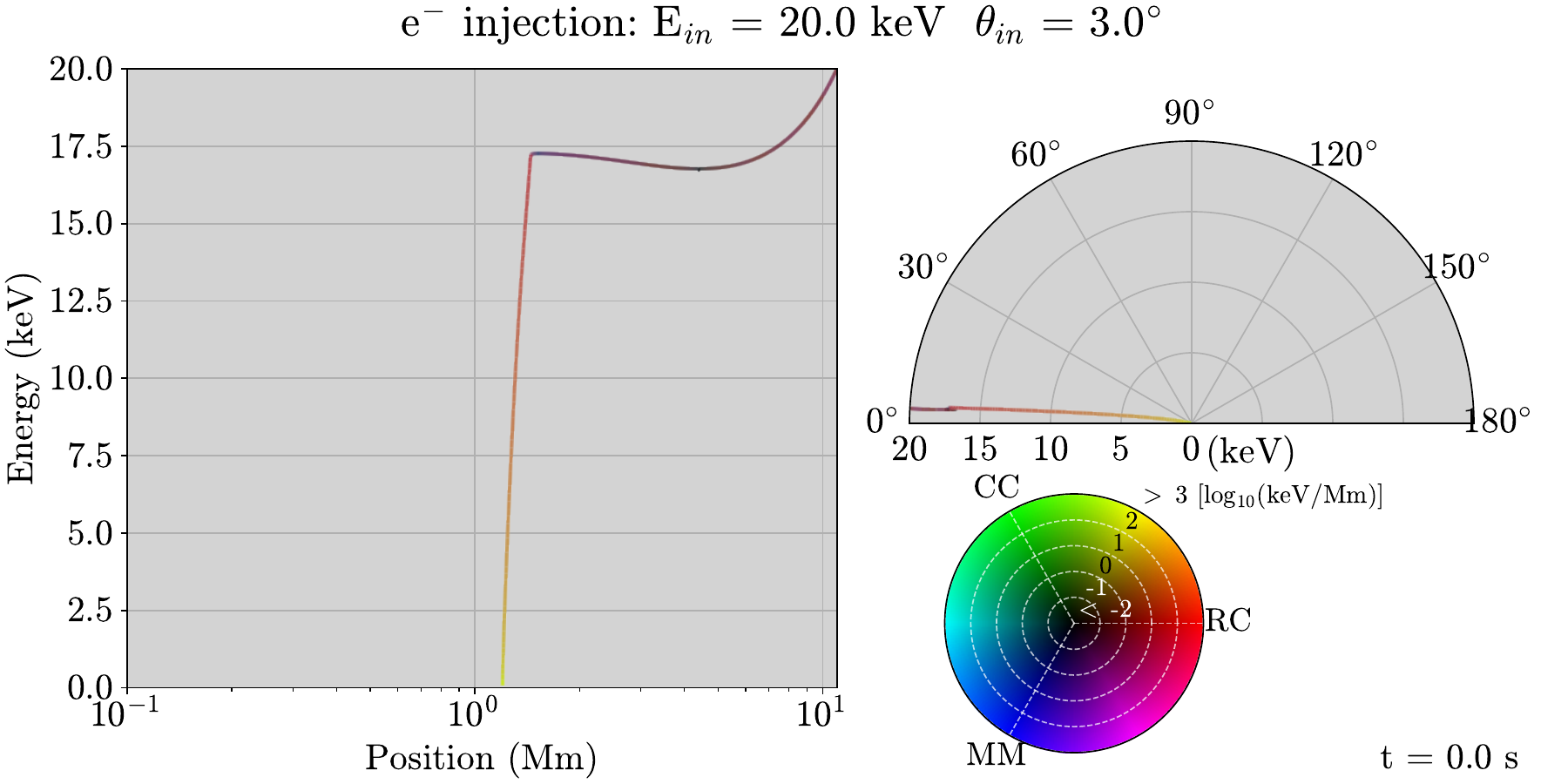}
    \plotone{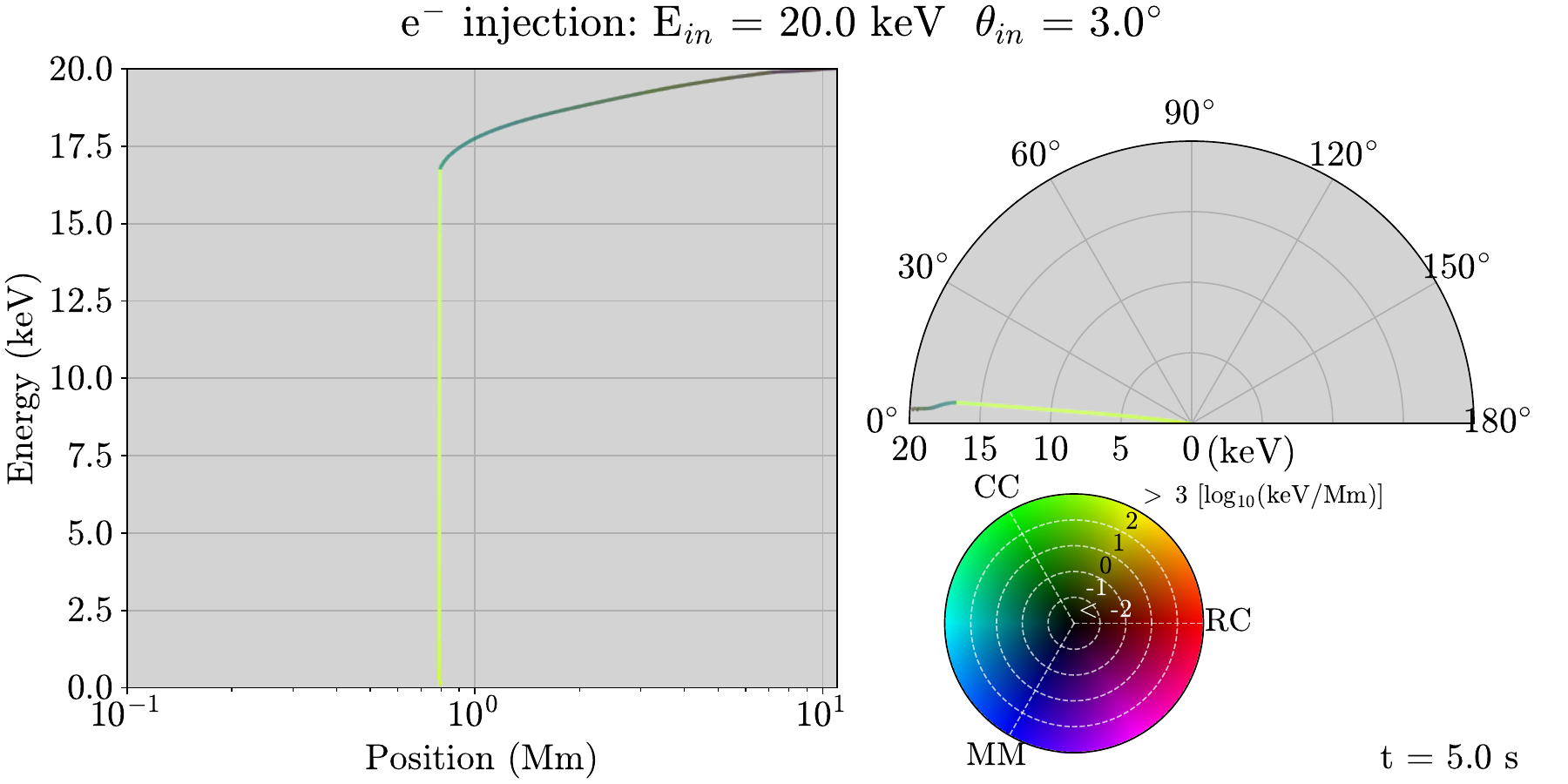}
    \end{interactive}
    \caption{Position vs energy (left plots in each panel) and pitch-angle vs energy (polar plots in each panel) for electrons injected with initial energy of 20 keV and at pitch-angle of 3$^{\circ}$. The color of the lines corresponds to the relative strengths of three dominant forces acting on the electrons. Red, green, and blue are the return current (RC), Coulomb collision (CC), and magnetic mirroring forces (MM), respectively, and the magnitude of the forces (in $\log_{10}$[keV/Mm]) is represented by the brightness, as indicated on the color wheel. An animation plotting these quantities for the full duration of electron injection ($t = 0$--99~s) is available online. Initially, the RC dominates energy loss in the corona, but by $t=0.5$~s CC dominates. As the simulation progresses, the electrons penetrate more deeply, finally reaching a position of $\sim600$~km. Snapshots showing $t=0$~s and $t=5$~s are shown in the top and bottom panels, respectively.}
    \label{fig:testpart_e_20_3}
\end{figure*}

\begin{figure*}
    \begin{interactive}{animation}{testpart_e_20keV_th45.mp4}
    \epsscale{0.9}
    \plotone{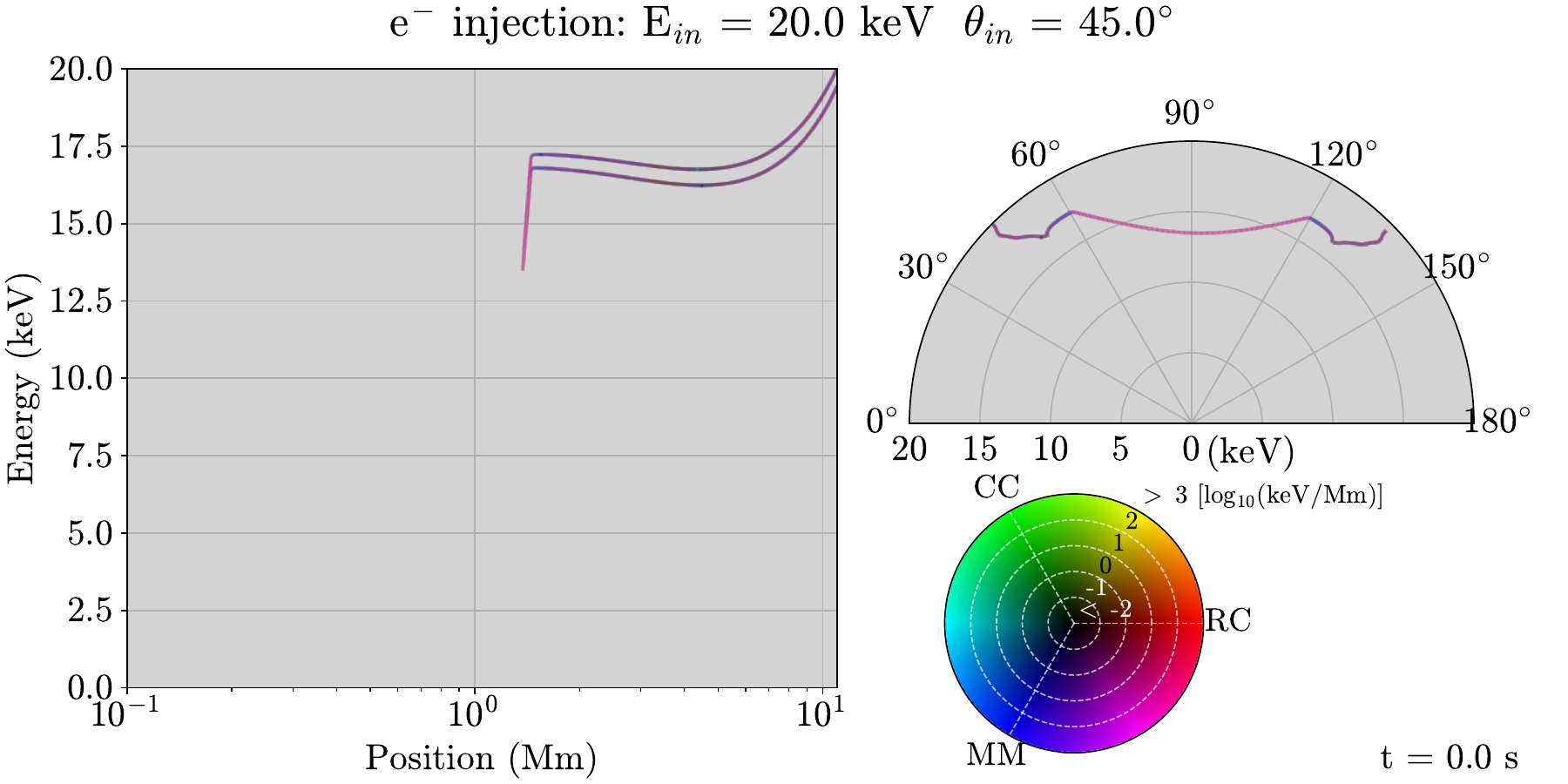}
    \plotone{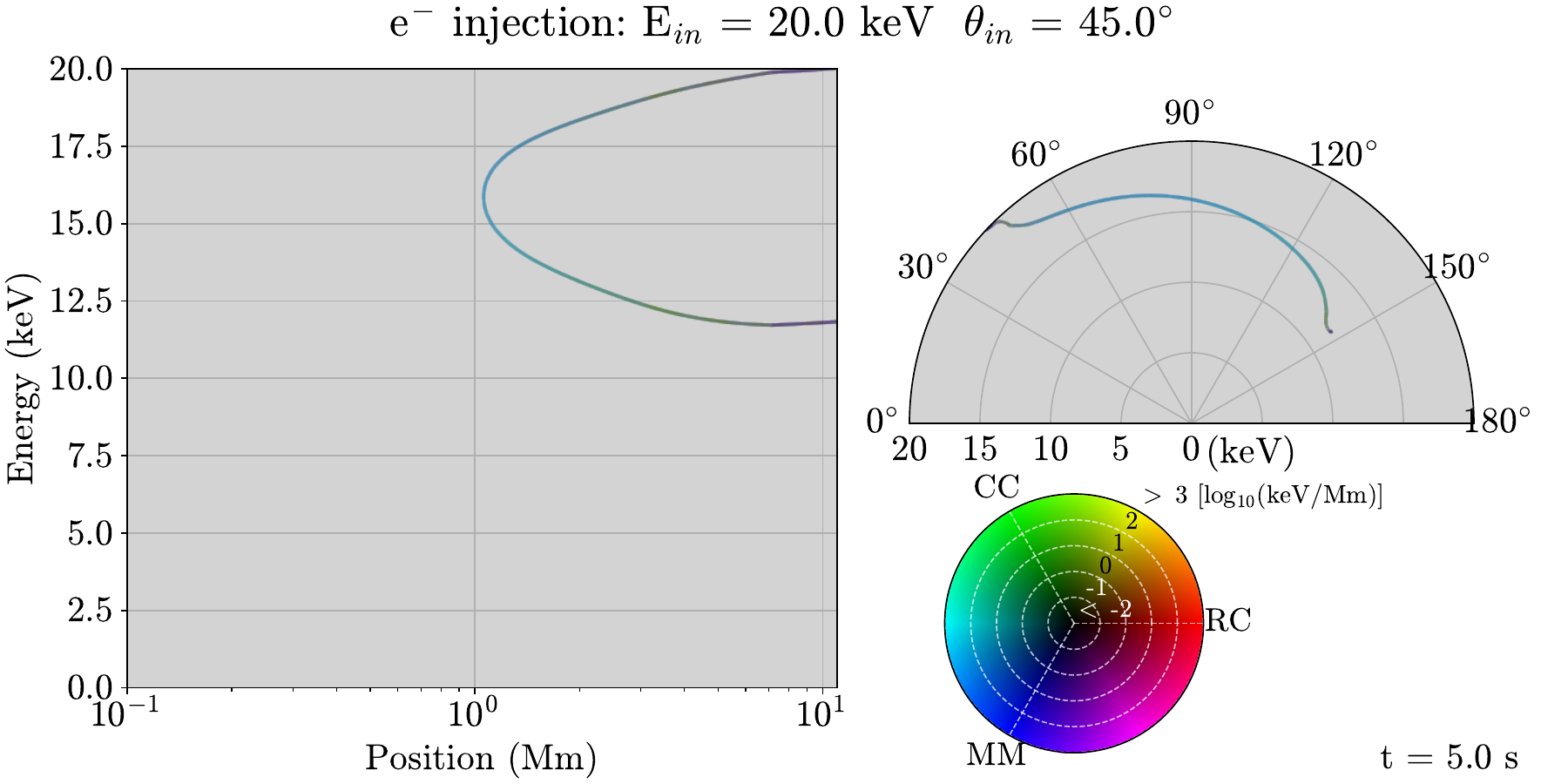}
    \end{interactive}
    \caption{Identical to Figure~\ref{fig:testpart_e_20_3} except for electrons injected with initial energy of 20 keV and at pitch-angle of $45^{\circ}$. An animation for the full duration of electron injection ($t = 0$--99~s) is available online. Initially strong RC and MM forces reverse the direction of the injected electrons, but by $t=0.5$~s the RC force has decreased, and the MM force dominates in reflecting the electrons. By $t=26$~s the corona has become dense enough for CC to stop the electrons. Snapshots showing $t=0$~s and $t=5$~s are shown in the top and bottom panels, respectively.}
    \label{fig:testpart_e_20_45}
\end{figure*}

The flux of downward-directed nonthermal electrons represents a large electrical current, which is neutralized by a return current composed of drifting ambient electrons, nonthermal electrons scattered back toward the looptop whether by Coulomb scattering or magnetic mirroring, and nonthermal protons traveling downward. More precisely, $J_{-} = -J_{+}$, where $J_{-}$ is the current density from downward-directed electrons and upward-directed protons, and $J_{+}$ is the current density from downward-directed protons, upward-directed electrons, and upward drifting ambient electrons. The fraction of each of these sources at several times is plotted in Figure~\ref{fig:rcfrac}. Initially, the return current exerts a strong force slowing the field-aligned component of the nonthermal electrons' velocities ($v_\parallel = |v|\mu$; in this section $\mu$ is the cosine of the pitch angle). As the nonthermal electrons enter the transition region, the temperature quickly drops, increasing the plasma resistivity and the return current electric force. For electrons with $\mu \approx 1$, the return current electric force slows them sufficiently to be thermalized (see Figure~\ref{fig:testpart_e_20_3} described in the paragraph below), but not so for electrons with $v_\perp = |v|\sqrt{1-\mu^2} \gg v_{th}$, where $v_{th}$ is the thermal velocity. These electrons are slowed until $v_\parallel$ reverses direction and they are re-accelerated back up the loop by the RC electric field, as shown in the top panel of Figure~\ref{fig:testpart_e_20_45} (also described in the paragraph below). In fact, initially these electrons are a dominant contributor to $J_{+}$. In this sense, the RC electric field tends toward self-regulation; an even stronger RC field would drive more nonthermal electrons back up the loop, which decreases the needed balancing drifting component and the corresponding electric field.

The trajectories of 20~keV electrons injected with nearly field-aligned velocities ($\theta = \arccos(\mu) = 3^{\circ}$) and with significant perpendicular components ($\theta = 45^{\circ}$) are shown in Figures~\ref{fig:testpart_e_20_3} and \ref{fig:testpart_e_20_45}. In these figures and their corresponding movies, each frame represents the full trajectory in that hydrodynamic snapshot. Then the subsequent frame tracks the trajectory of a newly injected particle on that snapshot etc. The relative strengths of three forces are encoded by the Red-Green-Blue color space. The red, green, blue color axes refer to the return current electric force (RC), Coulomb collisional force (CC), and magnetic mirroring force (MM), respectively. So that, for example, red lines mean that the RC force is large; yellow means the RC and CC forces are both large; magenta means the RC and MM forces are large, etc. Since the initial atmosphere is cool, the plasma resistivity is large and Ohmic losses from the RC electric field dominate. The electrons initially lose about 2.5 keV in their transport through the corona. In the low pitch-angle case (top panel of Figure~\ref{fig:testpart_e_20_3}), as the electrons travel through the transition region they encounter cooler and denser plasma, causing higher RC and CC losses. Therefore they lose sufficient energy to be thermalized at about 1.1~Mm, which is deep in the chromosphere. Interestingly, for the $\theta = 45^{\circ}$ case, as mentioned above $v_\perp \gg v_{th}$ so they are not thermalized by the RC. Instead, initially the RC electric field turns them around and re-accelerates them back up the loop (top panel of Figure~\ref{fig:testpart_e_20_45}). At a position of about 4.5~Mm the flux of protons exceeds that of electrons and the return current switches direction. This can be seen as a small negative component of the RC between 4.5~Mm and 1.6~Mm in the top left panel ($t=0$~s) of Figure~\ref{fig:rcfrac}. A resulting small RC acceleration of electrons over the same range is evident (top left panels of Figures~\ref{fig:testpart_e_20_3} and \ref{fig:testpart_e_20_45}). By $t = 0.1$~s, MM is the dominant force reflecting the electrons, but the RC continues to re-accelerate reversed electrons for approximately the first 0.5 s (first frames of animation associated with Figure~\ref{fig:testpart_e_20_45}).

At later times (e.g., $t = 5$~s in the bottom panels of Figures~\ref{fig:testpart_e_20_3} and \ref{fig:testpart_e_20_45}) the corona has heated and the transition region has been pushed deeper. In the low pitch-angle case, the RC and CC losses dominate, resulting in thermalization of electrons deep into the compressed chromosphere ($0.8$~Mm). The transition region and chromosphere have been pushed below the converging magnetic mirror at about 1~Mm. For the high pitch-angle case, the electrons encounter a sufficiently strong MM force to reflect them back up the loop. However, chromospheric evaporation has increased the density, resulting in 9~keV losses (at $t = 5$~s) due to CC during the electrons' transport through the corona. As chromospheric evaporation continues to build density in the corona, eventually ($t > 25$~s) the CC losses are sufficient to thermalize the electrons before they reach the looptop. This is evident in the animated version of Figure~\ref{fig:testpart_e_20_45} available online.

\subsubsection{Continuum emission\label{sec:cont}}

\begin{figure}
    \plottwo{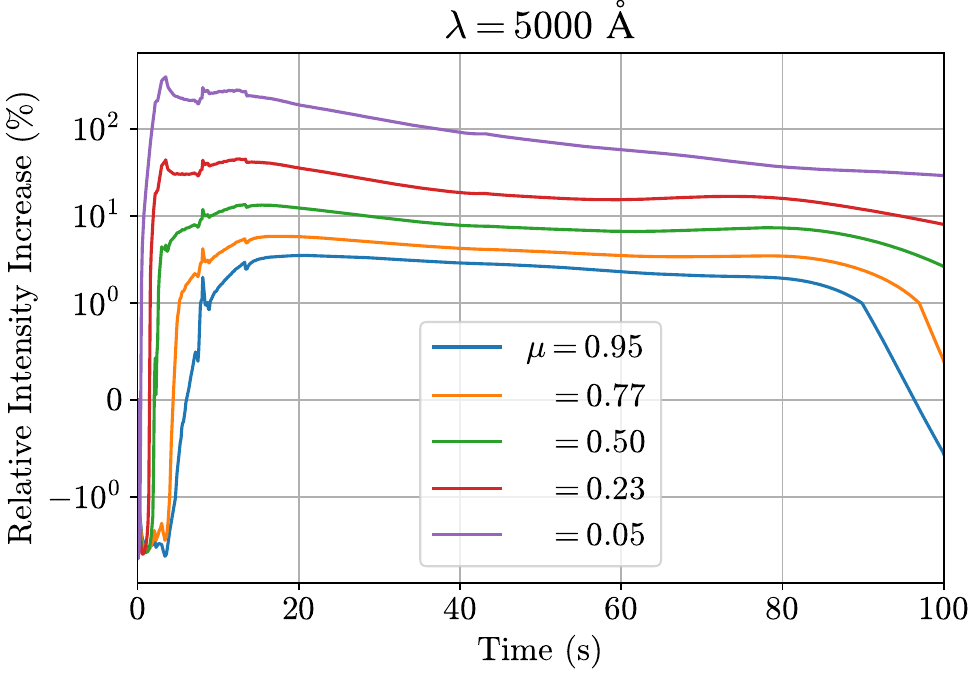}{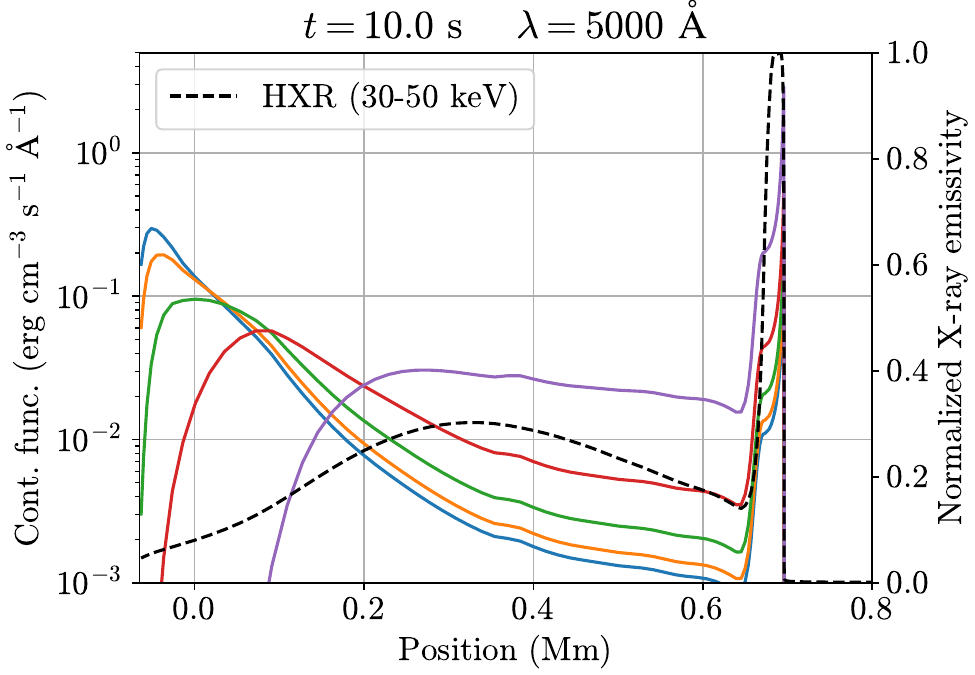}
    \caption{(Left) Relative increase in continuum vs time at $\lambda = 5000$~\AA\ for five viewing angles, where $\mu$ is the cosine of the viewing angle. (Right) Contribution integrals vs position for $\lambda=5000$~\AA\ at $t=10$~s for the same five viewing angles. The black dashed line shows normalized hard X-ray (30--50~keV) emissivity as a function of position.}
    \label{fig:wlcont}
\end{figure}

Flare-enhanced white-light (WL) continuum is potentially associated with deep atmospheric heating, perhaps via impacting ion beams \citep{2025ApJ...988...74G}. \radyn\ predicts continuum emission by solving the radiative transfer equation (see Section~\ref{sec:radynfp}) including sources of opacity important to the formation of the continuum, such as H$^{-}$ free-free and hydrogen, helium, and H$^{-}$ bound-free edges. We show the predicted continuum emission at $\lambda = 5000$~\AA\ for five viewing angles in the left panel of Figure~\ref{fig:wlcont}. In each case, the continuum initially decreases by about 5\% due to the rapidly increasing nonthermal collisional excitation of hydrogen, which increases the absorption of photospheric emission \citep{1990A&A...233..577H,1999ApJ...521..906A}. The increased opacity raises the optical depth $\tau = 1$ surface to higher, cooler layers. The time duration of the continuum decrease depends strongly on the viewing angle lasting $0.4$~s near the limb ($\mu = 0.05$; in this section $\mu$ is the cosine of the viewing angle) and $6.0$~s near disk center ($\mu = 0.95$). Short lived dimmings have been observed during flares in the IR continuum \citep{2003A&A...403.1151D} and in the He~I 10830~\AA\ line \citep{2016ApJ...819...89X}, with modeling suggesting that this is indeed a result of nonthermal collisional ionization \citep{2021ApJ...912..153K,2024ApJ...970...21K}. After the initial dimming, the continuum quickly rises, peaking at larger contrasts for increasing viewing angle. We find the increased continuum is primarily due to optically thin hydrogen bound-free emission forming at the base of the very steep transition region, consistent with \citet{2014ApJ...794L..23H,2017ApJ...847...48H} and \citet{2018A&A...620A.183J}. Since the transition region is very thin when viewed along the loop axis at disk center, its volume is relatively small and the continuum increase peaks at a modest 5\%. However, when viewed at the limb, the line-of-sight is nearly perpendicular to the loop, passing through the loop's cross-section with a larger emitting volume. In this case, the increase peaks at nearly 400\%. Of course, larger contrasts are also expected at the limb due to limb darkening.

To determine the origin of the continuum emission in the flaring loop, we consider the contribution integral $I = \int (j(z)/\mu) e^{-\tau(z)/\mu} dz$, where $j(z)$ and $\tau(z)$ are the emissivity and opacity, respectively. The right panel of Figure~\ref{fig:wlcont} shows the integrand of the contribution integral as a function of position for the same five viewing angles. At disk center most continuum emission originates in the photosphere, with a small contribution from hydrogen bound-free emission at the base of the transition region. The increased opacity is insufficient to significantly move the $\tau = 1$ surface upward. However, at the limb this situation is reversed. The extra path length along this line-of-sight increases the opacity sufficiently to stretch the region with $\tau$ near unity to higher altitude and over an extended range. Moreover, for this line-of-sight the bulk of the WL continuum enhancement originates from hydrogen bound-free emission forming at the base of flaring transition region, which has been compressed and pushed lower (690~km at $t=10$~s). The viewing angle at which the WL continuum crosses from dominantly photospheric to chromospheric is at $\mu = 0.06$.

A few studies have analyzed WL flares observed close to the limb to determine the altitude of the flaring hard X-ray (HXR) emission above the photosphere \citep{2012ApJ...753L..26M,2015ApJ...802...19K} . For comparison, we have plotted the hard X-ray emissivity in the energy range 30--50~keV (black dashed line) in the right panel of Figure~\ref{fig:wlcont}. The X-ray emissivity was calculated from our model using the bremsstrahlung cross-section \citep{1997A&A...326..417H}, including the Elwert multiplicative factor, with a mean atomic number $Z_a =1.2$. As expected the bulk of the hard X-rays are emitted where beam heating is peaked, at the base of the transition region, co-spatial with the WL enhancement at an altitude of 690 km above the surface photospheric surface and approximately 400 km above the limb. These values compare well with those measured by \citet{2015ApJ...802...19K}, who found WL enhancements and HXR emissions at ``altitudes around 800 km above the photosphere or 300-450 km above the limb height.'' They conclude that ``To be compatible with the standard thick target model, these rather low altitudes require very low ambient densities within the flare footpoints.'' However, our study provides another interpretation. We have found momentum deposition from impacting nonthermal particles and pressure gradients from their resultant heating compress the lower atmosphere and push the flaring transition region and chromosphere to relatively low altitudes. Hence we find these altitudes consistent with the thick-target model even at typical ambient chromospheric densities.

\section{Conclusions and Summary} \label{sec:conc}
We have described a new framework for constructing comprehensive solar flare models. Flare dynamics span many scales, and no current model is able to capture all processes at their native scales. Our framework addresses this challenge by chaining outputs from three well-tested flare modeling codes, each focusing on different aspects and dynamics of the overall flare evolution.
\begin{enumerate}
    \item We used the global-scale MHD code \arms\ to predict the formation of a current sheet from a sheared magnetic arcade in the breakout configuration. The field undergoes magnetic reconnection in the current sheet, producing the newly reconnected loops that comprise the flare arcade. We selected a newly reconnected field line and tracked the evolution of its morphology and physical properties from onset to impact at the top of the arcade. The final state provided input for subsequent calculations with \kglobal\ and \radynfp.
    \item We used the local plasma $\beta$ and Alfv\'en speed in the selected loop as input to the \kglobal\ code, which predicts distributions of energized electrons and ions undergoing Fermi acceleration in the reconnecting current sheet. 
    \item We initialized the flare-loop code \radynfp\ with the hydrodynamic state of the evolved \arms\ field line. We injected the distributions predicted by \kglobal\ into that flare loop and tracked its radiative hydrodynamic response. The \fpc\ component follows the transport of the injected energized particles and models their interactions with the ambient plasma through Coulomb collisions, return currents, and magnetic mirroring. Their collisions transfer energy and momentum to the ambient plasma, which serve as source terms in \radyn\ to predict the flare loop’s hydrodynamic evolution and radiative output.
\end{enumerate} 

Using our comprehensive framework, we investigated in detail the hydrodynamic evolution of a flare loop. The loop model predicted chromospheric evaporation upflows that were fast initially, with speeds of 1.7~Mm~s$^{-1}$, because of the initially low density, but decreased to 160~\kms\ as the loop filled with ablated, heated plasma. Chromospheric condensation downflows were predicted with velocities in the range of 10--30 \kms. The crossover from upflow to downflow occurred in plasma with temperatures in the range of 1--2~MK. Chromospheric condensations compressed the transition region and chromospheric footpoints, pushing this interface to approximately 0.7~Mm above the photospheric surface. This altitude is consistent with those inferred from X-ray observations at the solar limb \citep{2015ApJ...802...19K}. 

The current density from injected nonthermal particles was balanced by a return current, and we investigated the contributions of nonthermal and drifting ambient plasma to the return current. We found that initially the return current was composed primarily of nonthermal electrons that had been re-accelerated upward by the return-current electric field, or reflected by a chromospheric magnetic mirror. At later times, the increasing coronal temperatures caused lower return-current electric fields, so the bulk of the return current was then comprised of drifting ambient electrons.

We investigated the relative effects of Coulomb collisions, return currents, and magnetic mirroring on nonthermal electrons injected with low and high pitch angles. We found that return current and Coulomb collisions thermalize low pitch-angle electrons in the footpoint. Higher pitch-angle electrons are initially reversed and re-accelerated by the return-current electric field. As the coronal temperature rises, this electric field weakens, and newly injected high-pitch-angle electrons penetrate into and are reflected by the magnetic mirror in the chromosphere.

We have made our simulations as realistic as possible, but we note the following limitations:
\begin{enumerate}
    \item Each of the constituent models has its own limiting assumptions, which are discussed in works cited in Section~\ref{sec:codes}. We summarize a few here. 
    \begin{enumerate}
    \item The \arms\ simulation does not include the stratified lower atmosphere or thermodynamic processes such as conduction, radiation, and heating. Alfv\'en waves generated by reconnection are visible in the \arms\ simulation, but are not propagated into \kglobal\ or \radynfp. 
    \item \kglobal\ assumes periodic boundary conditions. \kglobal\ models Fermi acceleration from contracting magnetic islands, but other acceleration mechanisms, such as stochastic acceleration or acceleration at a termination shock, are not included in the model.
    \item \radynfp\ assumes a loop that is symmetric about the apex. In addition, the decay time scale of particle injection is assumed to equal the bursty reconnection time scale, which we take to be 20~s. In transporting injected nonthermal particles, \fpc\ finds time-independent solutions on individual \radyn\ hydrodynamic snapshots. This implicitly assumes that the nonthermal particles' times-of-flight are much faster than the hydrodynamic evolution time scale, which, for nonthermal protons, is only true after the evaporation front reaches the loop apex. 
\end{enumerate} 
\item Limitations of the entire framework include:
\begin{enumerate}
    \item Our framework does not account for bi-directional feedback between the chained models. That is, information flows from \arms\ to \kglobal\ to \radynfp\ but not back up the chain. This approach is justifiable for several reasons. Because \kglobal's spatial and temporal scales are much smaller than \arms's, \kglobal\ reaches a quasi-steady state essentially within each \arms\ snapshot. Moreover, the \arms\ loop provided to \radynfp\ is nearly fully evolved to MHD equilibrium. The intense heating from injected flare-accelerated particles drives flows that are constrained magnetically along that loop and will have little effect on the evolution of neighboring flare loops. 
    \item The framework does not yet model the temporal evolution of the entire system, only the plasma response to a single episode of beam injection into a snapshot of a newly reconnected loop in \arms.   
    \item We have applied our flare modeling framework to a single loop produced from a 2.5D MHD \arms\ simulation. Of course, real flares are intrinsically 3D and are comprised of many loops organized into flare arcades. 
 \end{enumerate} 
 
 The framework as described here is an initial step toward more comprehensive simulation of the complex, multiscale flaring phenomenon. 
 In future work, we will utilize 3D \arms\ simulations scaled to observed active-region dimensions, and apply our novel framework to construct multi-dimensional flare arcades comprised of hundreds of individual loops using a technique similar to \citet{2020ApJ...900...18K}. Ingesting multiple \arms\ snapshots into the framework will allow us to model the evolution of the system self-consistently from flare onset to decay phase. We will also perform parameter studies to investigate the influence of key variables, such as the beam injection profile, which will provide valuable insight into the wide range of observed flare responses and diagnostic signatures. 
\end{enumerate}

\begin{acknowledgements}
The authors acknowledge funding from NASA's Heliophysics Theory, Modeling and Simulations program. GSK acknowledges funding through NASA Grant 80NSSC24K1443, and later support from a Royal Society University Research Fellowship URF\textbackslash R1\textbackslash 251160.
JTD and MS acknowledge funding through NASA Grant 80NSSC24K0608. SEG acknowledges funding through NASA Grant 80NSSC24K0590. This research was partially supported by the International Space Science Institute (ISSI) in Bern, through ISSI International Team project \#25-630 ``Next-Generation Solar Flare Modelling: Bridging The Multi-Dimensional Gap.''
\end{acknowledgements}

\clearpage

\bibliography{htms_1}{}
\bibliographystyle{aasjournalv7.1}

\end{document}